\documentclass[12pt]{article}

\usepackage{newtxtext,newtxmath}
\usepackage{multirow}

\usepackage{caption}
\usepackage{scicite}

\usepackage{url}

\usepackage{hyperref}

\usepackage{graphicx}
\usepackage{placeins}

\usepackage[letterpaper,margin=1in]{geometry}

\renewenvironment{abstract}
	{\quotation}
	{\endquotation}

\date{}

\makeatletter
\renewcommand{\fnum@figure}{\textbf{Figure \thefigure}}
\renewcommand{\fnum@table}{\textbf{Table \thetable}}
\makeatother

\def\scititle{
	Static heterogeneity generates apparent universality in first-passage bursty dynamics
}
\title{\bfseries \boldmath \scititle}

\author{
	M. M\o ller$^{1\dagger}$,
	P. Rahe$^{2}$,
	S. Ghaderzadeh$^{3}$,
    E. Besley$^{3}$,
    P. Moriarty$^{1\ast}$\and
	\small$^{1}$School of Physics \& Astronomy, University of Nottingham, Nottingham NG7 2RD, UK\and
	\small$^{2}$ Fachbereich Physik, Universit\"at Osnabr\"uck, Barbarastra\ss e 7, 49076 Osnabr\"uck, Germany\and
    \small$^{3}$School of Chemistry, University of Nottingham, Nottingham NG7 2RD, UK\\
	\small$^\ast$Corresponding author. Email: philip.moriarty@nottingham.ac.uk\\
	\small$^\dagger$ Current address: National Institute of Chemical Physics and Biophysics, Tallinn, Estonia.
}

\begin{document} 

\maketitle

\begin{abstract} \bfseries \boldmath
\vspace{-4mm} \noindent Processes involving bursts of activity separated by quiescent periods occur across diverse systems and scales. In human dynamics, these phenomena have been described by power-law inter-event time distributions, \(P(t)\sim t^{-\alpha}\), with putative universality classes \(\alpha=1\) and \(\alpha=\frac{3}{2}\) having been proposed. Whether the observed $\alpha = 1$ scaling reflects intrinsic scale-free dynamics or instead emerges from heterogeneous underlying rates has been debated at length. We address this question in a canonical physical system for first-passage dynamics: two-dimensional molecular diffusion detected by the tip of a scanning tunnelling microscope. The resulting inter-pulse time distributions exhibit the same apparent truncated power-law form reported for human activities such as email communication, web browsing, and library loans. Maximum-likelihood estimation and model comparison decisively favor a \mbox{Kohlrausch-Williams-Watts--tempered} power law, \(P(t)\propto t^{-\alpha}\exp[-(t/t_c)^\beta]\), with \(\alpha \sim 1\). Kinetic Monte Carlo simulations reproduce this behavior, showing that the apparent \(\alpha \sim 1\) scaling is confined to a finite time window and arises from tip-induced spatial heterogeneity, not scale invariance.
\end{abstract}

\section*{Introduction}
\noindent On first consideration it may seem reductionist in the extreme, and misguided at best, to attempt to frame human dynamics in terms of interatomic or intermolecular interactions. There are, however, very many surprising examples of collective social behaviour being captured more than adequately by statistical physics models first used to understand the behaviour of ensembles of atoms or molecules. Sociophysics\cite{Schweitzer2018} and econophysics\cite{Bouchaud2023} take staples of statistical mechanics -- including, in particular, the ubiquitous Ising model\cite{Macy2024} -- and retask/retool them for the analysis of societal interactions. In line with Philip W Anderson's famed ``more is different" maxim\cite{Anderson1972}, fascinating collective behaviour not expected at the individual level emerges from interparticle/inter-human interactions. In many cases, and arguably somewhat disconcertingly, humans can indeed be reduced to simple hard sphere/hard disk systems, Ising (and Ising-like) models, and/or cellular automata, whether it be in terms of the distribution of the speeds of attendees at music concerts\cite{Silverberg2013}, bumper car (``dodgem'') dynamics\cite{Buendía2017}, or rather more weighty societal phenomena such as the evolution of language\cite{Patriarca2023}, political polarisation\cite{Levin2021}, and xenophobia\cite{Barreira2013}. \\

\noindent Since the publication of a highly influential paper by Barab\'asi twenty-one years ago\cite{Barabasi2005}, an especially active sub-field of sociophysics has been the area often described as ``bursty'' human dynamics. The core argument of Barab\'asi's original 2005 work -- and numerous subsequent publications, see \cite{Karsai2018} for a comprehensive review -- is that a wide range of human activity patterns (including email and traditional ``snail mail'' communication, web browsing, and library visits) are not of purely Poisson, i.e. totally uncorrelated, character, as had been assumed for essentially a century or more. Instead, these communication patterns comprise bursts of activity separated by periods of inactivity, giving rise to strong deparatures from Poissonian behaviour; the interevent, or interpeak, times ($t_{\mathrm{IP}}$) follow heavy-tailed probability distributions that, according to Barab\'asi\cite{Barabasi2005}, are best described by a power law, $P(t_{\mathrm{IP}}) \sim t_{\mathrm{IP}}^{-\alpha}$. (This claim was, however, later revised to incorporate an exponential truncation of the power law\cite{Vazquez2005, Vazquez2006}, not least because a probability density function following a pure $t^{-1}$ dependence cannot be normalised.) As such, the dynamics were described as scale-free -- a particularly important claim in the evolution of the field of network science. Barab\'asi's proposal of power law dependence and the concomitant universality have, however, been the subject of extensive debate\cite{Johansen2006, Stouffer2005,Barabasi2005Reply,Jacomy2020, Broido2019,FoxKeller2005}, with Fox-Keller\cite{FoxKeller2005} providing a particularly insightful critique.\\

\noindent If, however, we put aside (for now) the various competing interpretations of the inter-event time distributions, it is nonetheless clear that an exceptionally broad variety of systems and phenomena, spanning many orders of magnitude in time and/or space, exhibit bursty behaviour. Earthquakes\cite{Corral2004}, neurons\cite{Beggs2012}, computer networks\cite{Karsai2011}, honeybees\cite{Gernat2018}, fruit flies\cite{Sorribes2011}, gene regulation\cite{Ghosh2015}, ion channel dynamics\cite{Hawkes1990}, housework chores\cite{Takeuchi2024}, mobile/cell phone communication\cite{Jiang2013}, electromigration\cite{Dalton2010} and financial markets\cite{Mathiesen2013} collectively comprise just a subset of systems exhibiting bursty behaviour. These are all complex systems and, as such, there has been a long-standing debate about the extent to which the bursty behaviour is driven by individual dynamics (such as the queuing models put forward by Barab\'asi, V\'azquez \textit{et al.}\cite{Barabasi2005, Vazquez2006}) versus interactions between various actors or components\cite{Choi2021,Lynn2019,Oliveira2009}, and/or the influence of external drivers (such as circadian/weekly cycles in the context of communication channels and social media usage)\cite{Malmgren2008,Malmgren2009, Ross2015}. \\

\noindent By focussing on a remarkably simple system, \textit{viz.} planar molecules diffusing in two dimensions (Fig.~1), we not only strip away the complexity inherent in other bursty systems but demonstrate that the so-called ``universality classes'' first put forward by Barab\'asi\cite{Barabasi2005} are paralleled at the molecular level.  Although this is not at all surprising for the $\alpha=\frac{3}{2}$ ``class'', given that it arises essentially from the dynamics of a random walk\cite{Redner2012, Vazquez2006}, the observation of the (apparent) $\alpha=1$ ``class'' is unexpected and, as we show, is a direct consequence of heterogeneity in the diffusive bias field.\\

\noindent Our measurement protocol — the detection of molecules diffusing under an STM tip \cite{Binnig1986,Lozano1995,Wang1998, Wander1998,Barth2000,Baud2003, Hahne2013,Hahne2014, HahneThesis, Schiel2022} — yields dynamics that bear at least a passing resemblance to the non-homogeneous Poissonian model introduced by Malmgren \textit{et al.}\cite{Malmgren2008, Malmgren2009} in the context of email communication. In their case, burstiness arises from the combination of Poisson statistics and a time-varying event rate driven by circadian and weekly activity cycles. In contrast, the bursty dynamics observed here reflect both the underlying system and the specifics of our detection method. The inhomogeneity is spatial in origin, imposed by the static potential landscape of the STM tip, but produces position-dependent variations in the molecular hopping rate and thus translates directly to temporal variability. (A key distinction must be drawn here, however, between an unbiased diffusion process itself, which is Markovian and underpinned by a Poisson distribution of hopping times, and trajectory-derived observables such as residence/interpeak/interevent time distributions. These distributions arise from diffusion pathways and thus reflect statistical memory of where the particle has been, even in the absence of bias. In other words, they depend on the global structure of the trajectory — not just the local hopping rates.) \\

\noindent In the following, we show that spatial heterogeneity alone is sufficient to generate the observed truncated power-law inter-event time distributions. In direct analogy to the periodic rate modulations invoked by Malmgren \textit{et al.}\cite{Malmgren2008}, we find that a single spatial Fourier mode alone is enough to drive a crossover to the experimentally observed $\alpha \sim 1$ scaling.

\section*{Results and Discussion}
\subsection*{Tracking molecular transit}
We monitor the diffusion of perylene-3,4,9,10-tetracarboxylic acid dianhydride (PTCDA) molecules on the Ag(110) surface using atom-tracking-mediated\cite{Pohl1988, Swartzentruber1996, Rahe2011} ultrahigh vacuum scanning tunnelling microscopy (STM) at room temperature \cite{methods}. As discussed previously, and extensively, for various PTCDA-on-metal systems\cite{Tautz2007, Mercurio2013, Ikonomov2008, Ikonomov2010a,Ikonomov2010b, Hahne2013, Bohringer1998, Levesque2016}, well-ordered molecular islands coexist with a 2D~gas of rapidly diffusing molecules at 300~K (Fig.~1(a)). For the ordered islands, the lengths of the PTCDA lattice basis vectors, namely 1.25~$\pm$~0.04~nm and 1.18~$\pm$~0.04~nm, and the angle between those vectors, \textit{viz.} 83$^{\circ}\pm 2^{\circ}$, are in very good agreement with the values determined by Seidel \textit{et al.}\cite{Seidel1997} for PTCDA/Ag(110) via both real space (STM) and reciprocal space (low energy electron diffraction) measurements. \\

\noindent Due to the high diffusion rate, the molecular gas appears as ``speckle'' noise in STM images in Fig.~1(a); the inset shows a simple schematic illustration of a freely diffusing molecule under the STM tip. Other tip states, including those that entirely block molecular diffusion, are also observed but are rejected for subsequent analyses other than to act as a control. (See Supporting~Text, figure~s3.)  If, instead of imaging, the tip is held at a fixed point over the molecular gas (using atom tracking protocols\cite{methods}) at a fixed bias (in this case, +0.5~V), strong temporal fluctuations in the tunnel current signal are observed (Fig.~1(b)) when a molecule diffuses under the tip, is resident for some time, and then diffuses away. As discussed at length by Hahne \textit{et al.}\cite{Hahne2013, Hahne2014, HahneThesis}, these fluctuations can be binarised by choosing a suitable threshold tunnel current\cite{methods} ($I_{\mathrm{th}}$ in Fig.~1(b); fig S1). (See Supplementary Text and fig. S2 for a discussion of the insensitivity of the distributions to the choice of $I_{\mathrm{th}}$ within reasonably wide limits.)  From the binarised signal, probability distributions for the peak widths, i.e. residence times ($t_{\mathrm{RT}}$), and peak separations, i.e. interpeak times ($t_{\mathrm{IP}}$), are readily extracted. (See~inset~to~Fig.~1(b)). 

\subsection*{Biased diffusion}

\noindent Although it is the analysis of the molecular interpeak time distributions (ITDs) that forms the primary focus of this paper, a discussion of the residence time distributions (RTDs) we have measured for PTCDA/Ag(110) is first warranted, both to put our results/system in the context of previous work\cite{Ikonomov2008,Ikonomov2010b,Hahne2014,HahneThesis,Schiel2022} and to provide key insights into the underpinning diffusive dynamics. In Fig.~2(a) we therefore show a set of raw, unbinned, RTDs as a function of setpoint tunnel current, $I_{\mathrm{SP}}$. (Choice of binning can substantially influence the visual representation of interpeak/interevent distributions and, at worst, introduce apparent, but entirely spurious, power law character to a log-log plot\cite{Blenn2016, FoxKeller2005}. We therefore prefer to present unbinned distributions for both experimental and simulated data throughout this paper.)\\

\noindent A truncated power law, $N(t_{\mathrm{RT}}) \propto t_{\mathrm{RT}}^{-\alpha} \exp(-t_{\mathrm{RT}}/t_{\mathrm{c}})$, where $N(t_{\mathrm{RT}}$) denotes the number of occurrences of a given (discretised) residence time and $t_{\mathrm{c}}$ is a characteristic time defining the rate of exponential decay, has been fitted to each of the measured RTDs (solid lines in Fig.~2(a)). Table~1 shows the value of $\alpha$ and $t_{\mathrm{c}}$ extracted from each fit, along with error bars determined via bootstrapping; the mean value of $\alpha$ across the six datasets is 1.53~$\pm$~0.03 (where the uncertainty here is $\pm1\sigma$). A dash–dotted line corresponding to a $-\frac{3}{2}$ gradient is included in Fig.~2(a) for reference. However, and as discussed at length by Clauset \textit{et al.}\cite{Clauset2009}, Fox-Keller\cite{FoxKeller2005}, and Blenn and Mieghem\cite{Blenn2016}, we again emphasise that apparent linear regions in log–log plots can too readily be misinterpreted: visually convincing ``power-law'' segments may be present even when the underlying scaling behaviour is, at best, ill-defined. (Our treatment of interpeak time distributions -- see following section -- is directly informed by this issue.) \\

\noindent Notwithstanding these caveats for now, there are nonetheless sound physical reasons underpinning the observation of an \mbox{$\alpha~\sim~\frac{3}{2}$} scaling regime, at least over a limited range of $t_{\mathrm{RT}}$, for the RTDs. As has previously been described in the context of the measurement of surface diffusion constants\cite{Hahne2013, HahneThesis, Ikonomov2010b, Schiel2022}, STM detection of diffusing molecules is essentially a variant of the classic first passage time problem\cite{Redner2012}. For diffusive motion, the characteristic distance explored by a random walker grows as $\ell(t) \sim t^{1/2}$, a hallmark of (unbiased) diffusion that is independent of microscopic details. In the presence of an absorbing boundary (i.e. the edge of the tip detection region), this diffusive scaling directly determines the survival probability (that is, the probability that the walker has not yet reached the boundary), leading to a long-time decay~\mbox{$S(t) \sim t^{-1/2}$}. The corresponding first-passage time distribution, i.e. the time derivative of the survival probability, therefore scales as $t^{-3/2}$ -- a universal consequence of diffusion-controlled first-passage processes. On this basis, the observation of \mbox{$\alpha \sim \frac{3}{2}$} scaling in Fig.~2(a) is unsurprising. \\

\noindent The comparatively extended $t^{-3/2}$ regime observed here does, however, differ from that produced by KMC simulations of the type described by Hahne\cite{Hahne2013}, where detection of diffusing molecules is via a single lattice cell. For that detection strategy, an extended $\frac{3}{2}$ scaling regime can be obtained only upon significantly increasing the effective molecular size (i.e. increasing the absorbing boundary length). We similarly found that incorporation of a larger effective detector area in KMC simulations was essential in order to reproduce the form of the experimental RTDs in Fig. 2(a)\cite{methods}. (See \textit{Kinetic Monte Carlo Simulations} below.) Moreover, Schiel, Rahe, and Maass\cite{Schiel2022} have analytically explored the influence of an attractive or repulsive harmonic potential (of varying effective stiffness, $k_{\mathrm{eff}}$) on residence time distributions; in the strongly interacting regime they find that the power law regime is washed out and only the exponential ``tail'' remains. This is not the case in~Fig.~2(a).\\

\noindent Our measurements and analyses differ from earlier approaches\cite{Hahne2013,Hahne2014,Ikonomov2010b,Lozano1995,Wander1998} in that our core goal is not to extract a diffusion constant from the asymptotic long-time limit of the residence-time and interpeak-time distributions (and/or the tunnel current autocorrelation function or power spectrum\cite{Lozano1995,Wang1998}). Instead, here the diffusive bias induced by the presence of the tip is not a perturbation to be minimised, but rather the central object of interest. Our focus is on the influence of a spatially inhomogeneous potential on the first-passage statistics, and hence the bursty dynamics, of a diffusing molecule detected within a fixed spatial region.\\

\noindent Variation of the tip--sample separation in the experimental measurements (via changes in the stabilisation setpoint tunnel current, $I_{\mathrm{SP}}$)  provides clear and compelling signatures of biased diffusion. First, although the functional form of the residence-time distribution does not change, an overall shift of the long-time tail to longer residence times is observed (Fig.~2(a) and Table~1). Higher values of the set-point tunnel current yield smaller tip-sample separations and thus strengthen the influence of the probe: $I_{\mathrm{t}} \propto \exp(-2\kappa d)$, where $I_\mathrm{t}$ is the tunnel current, $\kappa$ is the inverse decay length, and $d$ is the tip-sample separation. From the logarithm of the setpoint tunnel current, and making the usual (albeit large) assumption that $\kappa \sim~1$\AA$^{-1}$, we can deduce that there is a sub-\AA ngstrom change ($\Delta z \sim 70$~pm) in tip-sample separation when $I_{\mathrm{SP}}$ is varied from 50~pA to 200~pA. \\

\noindent Despite the shift in the tail of the RTDs, modifying the tip height via changes in $I_{\mathrm{SP}}$ leads to a relatively modest variation in the mean residence time, $\langle t_{\mathrm{RT}}\rangle$ (Fig. 2(b)), which increases from $\sim 40~\mu\text{s}$ to $\sim 70~\mu\text{s}$. This is unsurprising -- the tail of the distribution comprises residence time values that, while relatively large, are measured with very low probability and thus overall make a small contribution to the value of $\langle t_{\mathrm{RT}}\rangle$. The behaviour of $\langle t_{\mathrm{IP}} \rangle$ with respect to changes in $I_{\mathrm{SP}}$ is significantly more pronounced (upper plot in Fig.~2(b)), decreasing more than threefold (from $\sim$ 1.4 ms to $\sim$ 0.4 ms) as the tip is moved towards the surface and following an approximate inverse dependence on tip-sample separation.\\

\noindent We can also readily determine the probability of a molecule being found in the tunnel junction, \begin{equation} P_{\mathrm{occ}}=\frac{\Sigma t_{\mathrm{RT}}}{{\Sigma t_{\mathrm{RT}}+ \Sigma t_{\mathrm{IP}}}},
\end{equation}
where the summations are over all values of $t_{\mathrm{RT}}$ and $t_{\mathrm{IP}}$. As shown in Fig. 2(c), $P_{\mathrm{occ}}$ varies from 3\% to 15\% as a function of tip-sample separation, highlighting again the significant influence of the STM probe on PTCDA diffusion on Ag(110), even for small, sub-\AA ngstrom changes in the width of the tunnel barrier. (The linear dependence of $P_{\mathrm{occ}}$ on $\ln(I_{\mathrm{SP}})$ seen in Fig.~2(c) follows from the inverse dependence observed for ($\tau_{\mathrm{IP}}$) (Supplementary Text). \\

\noindent Ikonomov \textit{et al.}\cite{Ikonomov2010b} observed a dependence of the mean residence time on the tip-sample separation but carefully chose to work in a regime of negligible probe-molecule interaction. Their aim was to measure molecular diffusion constants in as non-invasive a fashion as possible. As noted above, we instead deliberately exploit the influence of the tip to explore how (tuneable) deviations from an unbiased random walk affect the distributions of $t_{\mathrm{RT}}$ -- and, more importantly, $t_{\mathrm{IP}}$ -- in the context of bursty dynamics. As such, the minimal variation in $\langle t_{\mathrm{RT}} \rangle$ as a function of tip-sample separation, combined with the pronounced dependence of $\langle t_{\mathrm{IP}} \rangle$ on $\ln(I_{\mathrm{SP}})$, would strongly suggest that a relatively long-ranged tip-sample interaction -- or, equivalently, bias field -- plays a key role in the diffusive dynamics.

\subsection*{Molecular inter-peak times and first-passage statistics}

At first glance, the interpeak time distributions (Fig.~3) for the PTCDA/Ag(110) system are apparently of the same broad functional form as the RTDs of Fig.~2, comprising a linear (or quasi-linear) region at short times followed by an exponential decay. One subtle difference is that a relatively broad shoulder for the least invasive tunnelling condition, $I_{\mathrm{SP}}$~=~50~pA, evolves into a sharper ``knee'' at $I_{\mathrm{SP}}$=200~pA, in line with the significant decrease in mean interpeak time (Fig. ~2(b)). At this juncture we also highlight that if the tip influence were solely of short-ranged character -- due to, say, the formation of a chemical bond or relatively localised van der Waals interaction between the apex and an underlying molecule -- we would not expect to see significant changes in the long-time tail of the ITDs, or in the value of $\langle t_{\mathrm{IP}}\rangle$,  as a function of tip height\cite{Hahne2013, Schiel2022}.\\

\noindent It is the regime below $\sim$~1~ms that is of most interest when it comes to comparison with the inter-event distributions of other ``bursty'' systems, including, for example, those described in \cite{Barabasi2005,Barabasi2005Reply,Vazquez2005, Vazquez2006, Malmgren2008, Malmgren2009, Karsai2011, Karsai2018}. In the inset to Fig.~3 we have expanded this region and also normalised the distributions (in the 2~$\times 10^{-5}$~s~$\leq\tau_{\mathrm{IP}} \leq 10^{-3}$~s range) so that they overlay each other. If we provisionally adopt the analysis strategy of Barabasi \textit{et al.}\cite{Barabasi2005, Vazquez2006}
(and many others\cite{Karsai2018}), i.e. search for (quasi-)linear behaviour on a log-log plot, then there would appear to be good evidence for ``$\alpha \sim 1$ scaling'' over approximately two orders of magnitude in $t_{\mathrm{IP}}$, i.e. comparable to the extent observed for email, web browsing, and library loan inter-event time distributions. \cite{Barabasi2005,Stouffer2005, Vazquez2006}. 

\subsection*{Maximum likelihood estimation} 
For all of the reasons emphasized by Clauset \textit{et al.}\cite{Clauset2009} and Shalizi\cite{Shalizi_powerlaw_2007}, however, visual assessment of linearity on a log-log plot (especially over a limited range) is not a robust method of detecting power law, or truncated power law, behaviour. We have therefore used the more statistically rigorous maximum likelihood estimation (MLE) framework\cite{Clauset2009, methods} to determine the functional form of the ITDs that is statistically most compatible with the measured data and to assess whether a 2D molecular gas can indeed be placed in the same putative ``universality class'' as systems that are up to ten orders of magnitude larger in both spatial and temporal scales\cite{Barabasi2005,Vazquez2006}.\\

\noindent A set of maximum-likelihood fits and information-criterion comparisons\cite{Akaike1974} was used to compare a variety of candidate distributions, spanning power law, log-normal, Weibull (stretched exponential), mixtures of exponentials, exponentially truncated power law, and a power law tempered by the Kohlrausch–Williams–Watts (KWW) function\cite{Kohlrausch1854,WilliamsWatts1970,Wu2016}, i.e.
\begin{equation}
N(t_{\mathrm{IP}}) \propto t_{\mathrm{IP}}^{-\alpha}\exp(-(t_{\mathrm{IP}}/t_{\mathrm{c}})^\beta)  
\end{equation}
\noindent where $\beta$ is a shape parameter. The KWW form was included as a flexible truncation model that accommodates both stretched ($\beta<1$) and compressed ($\beta>1$) exponential decay, and, more pertinently, because it is widely used to describe relaxation in heterogeneous dynamical systems\cite{Phillips1996, Laherrere1998, Wu2016, Chen2003}. Table~2 summarises the results of the MLE tests for the $I_{\mathrm{SP}}=150~\text{pA}$ data; the KWW-tempered power law form overwhelmingly provides the best description of the measured interpeak time distribution. This is true for all other measured ITDs, except for the $I_\mathrm{SP}=200~\text{pA}$ case where an exponentially truncated power law with $\beta=1$ is a (marginally) better description of the distribution. (See Table~3 and table~S1). In Fig.~4(a) we show the fit of the ITD for $I_{\mathrm{SP}}=150$ pA to the KWW-tempered power law.  Table~3 lists the values of $\alpha$, $\beta$, and $t_{\mathrm{c}}$ extracted from each experimental ITD via an MLE approach, along with the 16\%-84\% confidence intervals. Averaged over all six datasets, $\alpha =0.94 \pm 0.04$.\\ 

\noindent From Table~3 it is clear that, as the tip approaches the surface, the value of both $\beta$ and $t_{\mathrm{c}}$ decreases. However, other than the $I_{\mathrm{SP}}=200~\text{pA}$ text, the value of $\beta$ remains larger than 1 throughout. The cut-off is therefore of compressed exponential\cite{Cipelletti2000,Cipelletti2005}, rather than stretched exponential\cite{Phillips1996}, form across the majority of the setpoint range. The dominant trend is the pronounced reduction of $t_{\mathrm{c}}$ from $\sim$~15~ms at 50~pA to $\sim$~2~ms at 200~pA, indicating that the long-time tail is cut off at progressively shorter times as the tip is moved closer to the surface. This is naturally explained by the strengthening of the tip-induced potential: a deeper van der Waals well -- together with, as will be discussed in the following sections, heterogeneity in the energy landscape that becomes more important relative to $k_{\mathrm{B}}T$ -- increasingly confines the molecule to shorter excursions and promotes faster returns to the detection region. Meanwhile, the reduction of $\beta$ from $\sim$ 2.18 towards $\sim 1$ implies that the cutoff becomes less sharply compressed and more nearly exponential, consistent with a growing diversity of escape and return pathways under stronger tip-induced heterogeneity. In this sense, $t_{\mathrm{c}}$ sets the crossover scale for first-passage dynamics, whereas $\beta$ controls the manner in which the cut-off is approached.

\subsection*{Kinetic Monte Carlo simulations} To elucidate the origin of the KWW form of the ITDs we turn to KMC simulations, complemented and informed by density functional theory (DFT) and insights from analytical analysis\cite{Hahne2013, HahneThesis, Redner2012}. As a starting point, we again consider the simplest canonical case: an unbiased diffusing molecule in two dimensions, for which detection within a fixed spatial region constitutes the standard textbook first-passage problem\cite{Redner2012,Hahne2013}. A detailed comparison between analytical theory and kinetic Monte Carlo simulations of the interpeak-time distributions for this unbiased reference case is provided in the Supplementary Text (fig~S4); here we summarise only the key points.\\ 

\noindent For a two-dimensional random walk returning to a detection region, and in common with the residence time distribution, the ITD exhibits a genuine $\alpha = 3/2$ power-law regime, reflecting the same universal first-passage statistics that govern the RTDs in Fig.~2. Importantly, this $t^{-3/2}$ regime is intrinsically scale free: in the absence of additional length or time scales, its extent is limited only by external constraints such as finite system size (and, in our case, the associated periodic boundary conditions). On an infinite lattice, the $t^{-3/2}$ regime ultimately crosses over, at asymptotically long times, to the marginal recurrence limit $\psi(t_{\mathrm{IP}})\propto [t_{\mathrm{IP}}\ln^2 (t_{\mathrm{IP}})]^{-1}$\cite{Redner2012, Hahne2013}. In practice, however, this crossover is pushed to increasingly long times as the system size is increased, and for a single walker the appearance of an exponential cutoff reflects only the finite spatial extent of the system and the influence of periodic boundary conditions, i.e. the onset of ergodic behaviour. (As Hahne \textit{et al.}\cite{Hahne2013} discuss, an exponential cut-off will also result from the contributions of other molecules when $N_{\mathrm{mol}}>1$. The fundamental origin is the same however: a loss of memory due to the resetting of first-passage statistics. See the Supplementary Text (fig. S10 and related discussion) for the results of KMC simulations involving multiple molecules.)\\

\noindent While marginal recurrence therefore provides, in principle, a route to $\alpha\sim 1$ scaling, it would occur only in the extreme long-time limit of an effectively unbounded system and cannot account for the experimentally observed behaviour at short and intermediate times. This motivates the introduction of an external, tip-induced potential, which, as we now discuss, underpins the form of the measured interpeak time distributions.

\subsection*{Shaping the landscape: Constrained hopping}
To reproduce the $\alpha \sim 1$ scaling seen in the experimentally-measured ITDs (Fig.~3), we incorporate an external potential due to the STM tip into the KMC simulations. The tip-molecule interaction is treated using what might best be described as a physically-informed phenomenological model. In the context of bursty dynamics, we stress, however, that the precise microscopic origin of this potential is not the central issue; rather, what matters is that the tip introduces an externally-imposed bias with, as we shall see, sufficient spatial heterogeneity to appropriately modify the first-passage statistics associated with molecular diffusion.  With this perspective in mind, the aim of the following discussion is not to definitively identify the functional form of the interaction potential (which, in the absence of accompanying atomic force microscopy (AFM), Kelvin-probe force microscopy (KPFM), and/or scanning quantum dot spectroscopy\cite{Wagner2019} measurements cannot be uniquely identified from STM data in any case). Rather, it is to establish empirically-motivated constraints that allow several possible mechanisms for diffusive bias in the tip-PTCDA-Ag(110) system to be ruled out or at least relegated to unlikely/lower order contributions. These considerations in turn inform the modelling and ``sculpting'' of the interaction potential in the KMC simulations.\\

\noindent In principle, the tip–molecule interaction could be assumed to arise from electrostatic contributions due to either a permanent PTCDA dipole moment and/or the molecular polarisability interacting with the spatially inhomogeneous electric field of the STM tip\cite{Stroscio1991,Tsong1975,Wang1998,Simpson2023,Matvija2017,Rozboril2019}. For the PTCDA/Ag(110) system, however, a substantial body of experimental and theoretical work indicates that charge is transferred from the Ag(110) surface into the lowest unoccupied molecular orbital (LUMO) of the molecule\cite{Bauer2012,Alkauskas2006}, giving rise to a dipole moment oriented toward the surface. Under positive sample bias (as is the case in our experiments), this dipole should therefore experience a repulsive interaction with the tip. If such a dipolar interaction dominated the tip–molecule coupling, the tail of the ITDs would shift toward longer interpeak times—and the RTDs toward shorter residence times—as the tip approaches the surface and the electric field increases. This behaviour is instead entirely opposite to that observed experimentally, where PTCDA molecules are clearly attracted toward the tip under positive sample bias.\\

\noindent At first sight this might suggest, as in the work of Wang~\textit{et~al.}\cite{Wang1998}, that it is the interaction of the polarisability tensor -- rather than the static dipole moment -- with the $\mathbf{E}$-field of the tip\cite{Stroscio1991} that  provides the dominant contribution. However, the out-of-plane component of the polarisability tensor for PTCDA, at least in the gas phase\cite{Tsiper2001}, appears much too small to account for the magnitude of the tip-induced potential inferred from the experiments. A more plausible origin of the long-range attraction is therefore the geometry-enhanced, integrated van der Waals interaction associated with the finite radius of curvature of the STM tip\cite{Israelachvili2010}, which gives rise to a potential with an inverse power-law dependence\cite{Olsen2012, Israelachvili2010}, $U\propto 1/\rho^3$, where $\rho$ is the radial distance of the molecular centre from the tip axis in the surface plane. (We note that the $1/\rho^3$ dependence is, in any case, identical to that employed by Wang \textit{et al.}\cite{Wang1998} for Monte Carlo simulations of tip-induced adatom diffusion. See \cite{methods} for more detail on the form of the potential.)\\ 

\noindent Density functional theory (DFT) calculations\cite{methods} were first used to determine a rough estimate of the magnitude of the local tip–molecule interaction. Both clean (i.e. bare silver) and PTCDA-terminated tips were considered in the DFT calculations (see inset to Fig.~4(a) for an example of the latter and the Supplementary Text for the former. There is a propensity for surface-adsorbed molecules to attach to the STM tip during imaging of PTCDA/Ag(110) (fig s3, Supplementary Text). These calculations provided an estimate of the interaction energy close to the tip apex ($\sim 5 - 6~k_{\mathrm{B}}T$) at a reference tip height of 0.7 nm, which set the energy scale for the tip-induced diffusive bias incorporated in the KMC simulations. The spatial form of the potential was then constructed phenomenologically as outlined above to represent the longer-range interaction associated with the finite radius of curvature of the STM tip. In practice, the effective tip radius used in the simulations ($R=30$~nm) was chosen such that the resulting diffusion dynamics reproduced the experimentally observed variations in interpeak-time scale as a function of tip-height. \\

\noindent In order to concurrently reproduce the $\frac{3}{2}$ scaling seen in the experimentally measured RTDs (Fig.~2 and Table~1), both the detector geometry and the form of the potential close to the tip centre (i.e. $\rho=0$) required careful consideration. The lower inset to Fig.~4(a) highlights the influence of detector area. With a sharp, single-lattice-cell detector, entry and exit events are effectively governed by single hopping steps, suppressing the continuous first-passage dynamics necessary for the emergence of a clear $\frac{3}{2}$ regime and resulting in the observation of only an exponential ``tail'' (lowermost RTD in the inset). In contrast, the finite, smoothly weighted detector employed here\cite{methods} introduces a diffuse detection boundary, allowing entry and exit to occur over a distribution of trajectories and times, thereby recovering the expected scaling (to a greater or lesser extent depending on detector area). In the KMC simulations we chose a geometry that represented a compromise between a physically realistic detector area ($\sim 2.6$~nm$^2$; i.e. an effective radius of $r \sim 0.9$~nm, comparable to the dimensions of a PTCDA molecule) and the ``visibility'' of the $\alpha \sim 3/2$ regime (see RTD plotted in green in the inset to Fig.~4(a)). (The total detection area is essentially a convolution of the detector function with the simulated molecular footprint.)\\

\noindent Similarly, high potential curvature at the tip centre introduces a significant inward drift, driving the molecule rapidly toward the apex and thereby suppressing the relatively unbiased first-passage dynamics required for $\frac{3}{2}$ scaling. The introduction of a flat-bottomed core (of comparable size to the detector area) locally suppresses this drift, restoring quasi-unbiased diffusion within the well and enabling $\frac{3}{2}$ scaling to emerge over the experimentally relevant time window. ``Flat-bottomed'' geometries somewhat akin to this, albeit over much longer length scales), have been observed in transmission electron microscope (TEM) studies of STM tips following their imaging of metal surfaces\cite{Garnaes1990,Zhang1996, Naitoh1996}. On shorter length scales, our DFT calculations identified a number of stable binding geometries for a tip-adsorbed PTCDA molecule (Supplementary Text). This also provides a plausible mechanism for ``flattening'' of the tip potential.

\subsection*{Heterogeneity as the origin of $\alpha \sim 1$ scaling}  

A central objective in designing and paramaterising the KMC simulation was the reproduction of the KWW-tempered power law form of the ITDs observed experimentally (Fig.~4(a)). To this end, a range of physically motivated descriptions of the tip-induced potential were explored, alongside systematic variation of key model ingredients, including the tip radius of curvature (and hence the spatial gradient), the local form of the potential near the tip apex,  detector geometry, diffusion anisotropy, and the number of molecules (Supplementary Text). While each of these factors influences the dynamics, none was sufficient to reproduce the experimentally observed ITDs. Only upon introducing spatial heterogeneity in the energy landscape (Fig.~4(b)) did the characteristic $\alpha \sim 1$ scaling behaviour and KWW-tempered form seen in Fig.~4(a) (see also Table~4 and Table~S1) emerge.\\ 

\noindent In line with standard practice, our STM tips were prepared by voltage pulsing and indentation into the Ag(110) substrate. As such, the presence of atomistic, nanoscopic and mesoscopic roughness at the tip apex -- and the concomitant inhomogeneity in the tip-molecule interaction potential -- is to be expected. Not only is heterogeneity of the type seen in Fig.~4(b) in line with the combined TEM-STM studies mentioned above\cite{Garnaes1990,Zhang1996, Naitoh1996}, but atomistic simulations of realistic tip structures have highlighted that the apex and near-apex region can have significant levels of roughness on atomic and nanometre length scales\cite{Ghasemi2008,Pou2009,Jarvis2013}. In this context, dispersion-force-related energy variations of order a few $k_{\mathrm{B}}T$, as shown in Fig.~4(b), are a natural consequence of realistic tip structures.\\ 

\noindent We model the roughness of the potential via a simple Fourier summation over a limited number of modes,
\begin{equation}
\begin{aligned}
U(\mathbf{\rho}) &= U_0(\mathbf{\rho})\left(1+\epsilon\phi(\mathbf{\rho})\right), \\
\phi(\mathbf{\rho}) &= \sum_{n=1}^{N} A_n \cos\!\left(q_{x,n} x + q_{y,n} y + \varphi_n\right),
\end{aligned}
\end{equation}
\noindent where $U(\mathbf{\rho})$ is the overall energy, $U_0(\mathbf{\rho})$ is due to the unroughened vdW background, $\epsilon$ is the magnitude of the roughness modulation (in units of $k_{\mathrm{B}}T$), $A_n$ is the amplitude of the $n^{\mathrm{th}}$ Fourier mode, $q_{x,n}$ and $q_{y,n}$ are the $n^{\mathrm{th}}$ wavevectors in the $x-$ and $y-$ directions, and $\varphi_n$ is a phase term. The values of $A_n$, $q_{x,n}$, $q_{y,n}$, and $\varphi_n$ were randomly assigned for each realisation of the roughness field. Similarly, phases, $\varphi_n$, were drawn independently from a uniform distribution on $[0,2\pi]$.
The wavevectors were also constructed from randomly-oriented directions $\theta_n \sim [0,2\pi]$ and magnitudes $q_n = 1/\lambda_n$, where the wavelengths $\lambda_n$ were sampled from a Gaussian distribution centred on a characteristic length scale $\lambda_{\mathrm{c}}$ nm with standard deviation $\lambda_{\mathrm{\sigma}}$. The Cartesian components were then given by $q_{x,n} = q_n \cos\theta_n$ and $q_{y,n} = q_n \sin\theta_n$. The amplitudes $A_n$ were drawn from a uniform bounded distribution and normalised such that the variance of $\phi(\mathbf{r})$ remained of order unity. This construction yielded an isotropic, band-limited roughness field characterised by a finite number of spatial modes.\\ 

\noindent Surprisingly, however, we found that a single spatial Fourier mode is already sufficient to induce the crossover from the diffusive first-return scaling $\alpha_{\mathrm{ITD}} \sim 3/2$ toward the experimentally observed $\alpha_{\mathrm{ITD}} \sim 1$ (inset to Fig.~4(c)). Additional modes merely introduce further spatial variability without altering the core underlying mechanism. In this respect, single molecule diffusion provides a direct spatial analogue of the non-homogeneous Poisson process discussed by Malmgren \textit{et al.}\cite{Malmgren2008,Malmgren2009} in the context of social dynamics. (We note that the unroughened vdW potential, $U(\rho)\propto1/\rho^3$, can itself be expressed as a Fourier expansion. However, being monotonic in $\rho$, it does not introduce the heterogeneous local trapping required to broaden the return-time distribution.)\\ 

\noindent Although the local scaling exponent is insensitive to the number of Fourier modes (and to $0.5~\text{nm} \leq\lambda_{\mathrm{c}}\leq 5~\text{nm}$, Supplementary Text and figs.~S11,~S12), the main panel of Fig.~4(c) highlights that the value of the roughness modulation, $\epsilon$, has a critical influence on $\alpha_{\mathrm{ITD}}$. We estimated $\alpha_{\mathrm{ITD}}$ using two complementary approaches: (i)~linear~regression on the log–log representation of the ITDs, as commonly employed in the bursty dynamics literature\cite{Vazquez2006, Karsai2018}, and (ii) maximum likelihood estimation (MLE)\cite{Clauset2009} for a power-law distribution constrained to the \mbox{$20~\mu$s~$\leq t_{\mathrm{IP}}\leq1$}~ms window of interest. While the regression-based estimates are systematically larger until a value of roughness modulation, $\epsilon$, of  $\sim 0.4~k_{\mathrm{B}}T$ is reached, both methods yield similar overall trends as a function of~$\epsilon$. When the roughness of the potential is switched off by setting $\epsilon$ to zero -- in other words, the simulation includes only a smooth vdW potential that falls off monotonically as $U(r) \sim 1/\rho^3$ -- the scaling exponent, $\alpha_{\mathrm{ITD}}$, is $\sim 3/2$, i.e. the value expected from the first-passage dynamics for a random walker. With increasing roughness modulation, the value of $\alpha_{\mathrm{ITD}}$ decreases smoothly towards the $\alpha \sim 1$ seen in experiment. (Although the global mean square displacement (MSD) as a function of time is of purely diffusive character, a local, detector-conditioned MSD measure shows subdiffusive behaviour\cite{Bouchaud2012,BouchaudGeorges1990} Supplementary Text; fig S13).

\section*{Conclusions}
Our PTCDA/Ag(110) STM experiment and associated KMC simulations could be said to constitute an archetypal model of first-passage dynamics: we simply monitor (biased) random walks in two dimensions. Yet, despite the apparent simplicity -- particularly in comparison to the wide range of systems discussed in the ``bursty dynamics'' literature\cite{Karsai2018} --  we encounter the same central challenge: distinguishing genuine power-law scaling from other heavy-tailed forms is far from trivial. In this sense, it remains “suspiciously easy”\cite{FoxKeller2005} to infer apparent scale-free dynamics even in systems governed by comparatively simple stochastic processes.\\

\noindent We find that queueing or priority-based dynamics\cite{Barabasi2005,Vazquez2005} are not required for an effective inter-event exponent close to unity ($\alpha_{\mathrm{ITD}} \sim 1$) to emerge for a biased random walk. Instead, spatial heterogeneity in the underlying (time-independent) energy landscape, combined with the Boltzmann dependence of thermally activated hopping, maps static disorder onto a broad distribution of transition rates, thereby generating temporal inhomogeneity.\\ 
 
\noindent Echoing a point made in the introduction, it is tempting to draw parallels with human communication and social dynamics\cite{Malmgren2008,Malmgren2009,Barabasi2005,Vazquez2005,Vazquez2006}, where heavy-tailed inter-event times have been proposed to arise from the influence of external temporal structure such as circadian/work cycles\cite{Malmgren2008,Malmgren2009,Chu-Shore2010}. Our results show that comparable scaling in a model first-passage problem can indeed emerge without invoking explicit prioritisation or memory. In short, an exponent $\alpha \sim 1$ need not imply decision-making dynamics, but can arise generically from heterogeneity in the rates governing otherwise simple stochastic processes. That the statistical spread in the value of $\alpha_{\mathrm{ITD}}$ over different roughness realisations (Supplementary Text; fig.~S14) is very similar to that observed by V\'{a}zquez, Barab\'{a}si \textit{et al.}\cite{Vazquez2006} for web browsing, email communication, and library visits is perhaps suggestive in this context.

\newpage

\begin{figure}
\centering
\includegraphics[width=0.98\linewidth]{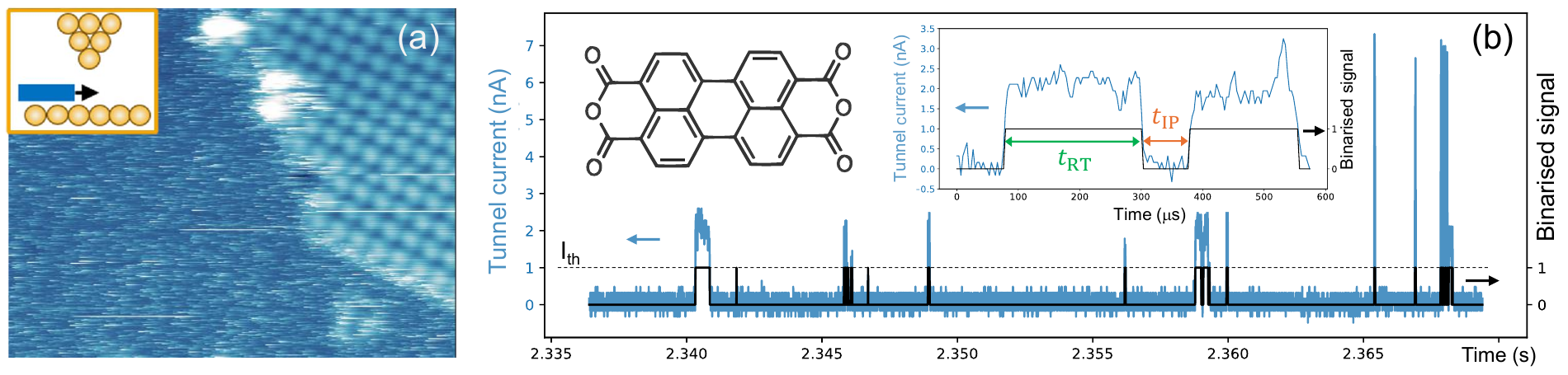}
\caption{\textbf{Bursty molecular dynamics.} \textbf{(a)} Room-temperature scanning tunneling microscope (STM) image (25.1~nm~$\times$~19.5~nm; sample bias:~+2V, setpoint tunnel current:~100~pA) of perylene-3,4,9,10-tetracarboxylic acid dianhydride (PTCDA) molecules adsorbed on Ag(110). A self-assembled, single-molecule-high PTCDA island occupies the right side of the image; the remainder of the image appears extremely noisy due to the rapid diffusion of molecules across the surface. For this particular tip state, PTCDA molecules diffuse under the tip (as shown schematically in the sketch in the inset). See Supplementary Text for examples of other tip states, including those that block molecular diffusion entirely. \textbf{(b)} Example of time series of tunnel current fluctuations at a fixed tip-sample bias of 0.5 V. $I_{\mathrm{th}}$ is the tunnel current threshold (in this case 1.0 nA) chosen for binarisation of the signal. The insets show (left) the planar structure of the PTCDA molecule, and (right) a zoom of part of the tunnel current signal for which the residence time and interpeak time, $t_{\mathrm{RT}}$ and $t_{\mathrm{IP}}$, respectively, have been highlighted.}    
\label{fig:1}
\end{figure}
\newpage 

\begin{figure}
\centering
\includegraphics[width=0.95\linewidth]{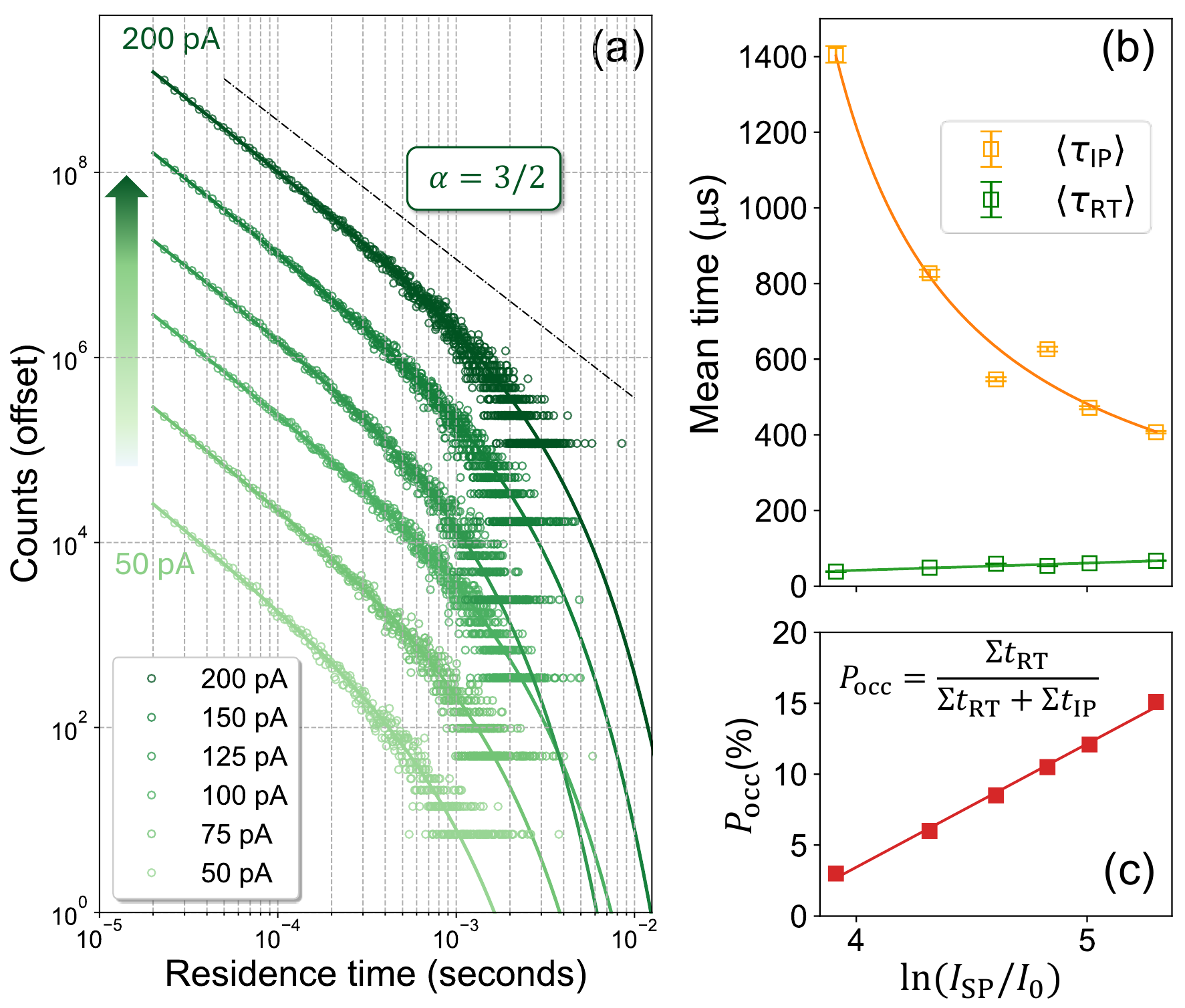}
\caption{\textbf{Residence time distributions for molecular diffusion.} \textbf{(a)}~Unbinned distributions of residence times for six different setpoint tunnel currents (at a fixed tip-sample bias of 0.5 V), which have been offset along the y-axis for clarity.  Higher set-point tunnel currents reduce the tip-sample separation. The dashed-dotted line is a guide to the eye that highlights the $\alpha=3/2$ gradient. Solid lines: each residence time distribution (RTD) is fit to an exponentially truncated power law of the form \mbox{$N(t_{\mathrm{RT}}) \propto t_{\mathrm{RT}}^{-\alpha} \exp(-t_{\mathrm{RT}}/t_c)$}. See Table 1 for a list of values of $\alpha$ and $t_{\mathrm{c}}$ determined from the fits and for associated uncertainties determined via a bootstrap approach. \textbf{(b)}~Variation of mean residence time, $\langle t_{\mathrm{RT}} \rangle$, and mean interpeak time,$\langle t_{\mathrm{IP}} \rangle$, as a function of the natural logarithm of the tunnel current set-point, $\ln(I_{\mathrm{SP}}/I_0)$ (where $I_0$ = 1 pA and is included only to formally ensure the argument of the logarithm is dimensionless), which is directly proportional to the tip-sample separation. Error bars on the mean values are determined from a bootstrapping approach (with 1000 samples). The solid orange line is a fit to the $\langle t_{\mathrm{IP}}\rangle$ data of the form 
$\langle t_{\mathrm{IP}} \rangle(d')=\langle t_{\mathrm{RT}} \rangle(\alpha + \beta d')^{-1}-\langle t_{\mathrm{RT}} \rangle$, where $d'=\ln(I_{\mathrm{SP}}/I_0)$ and $\alpha$ and $\beta$ are constants. (See Supplementary Text for the rationale for using this fitting function.) \textbf{(c)}~Probability of a molecule occupying the junction, $P_{\mathrm{occ}}$, as a function of $d'$. Error bars are smaller than the data marker size.}    
\label{fig:2}
\end{figure}
\newpage

\begin{figure}
\centering
\includegraphics[width=0.8\linewidth]{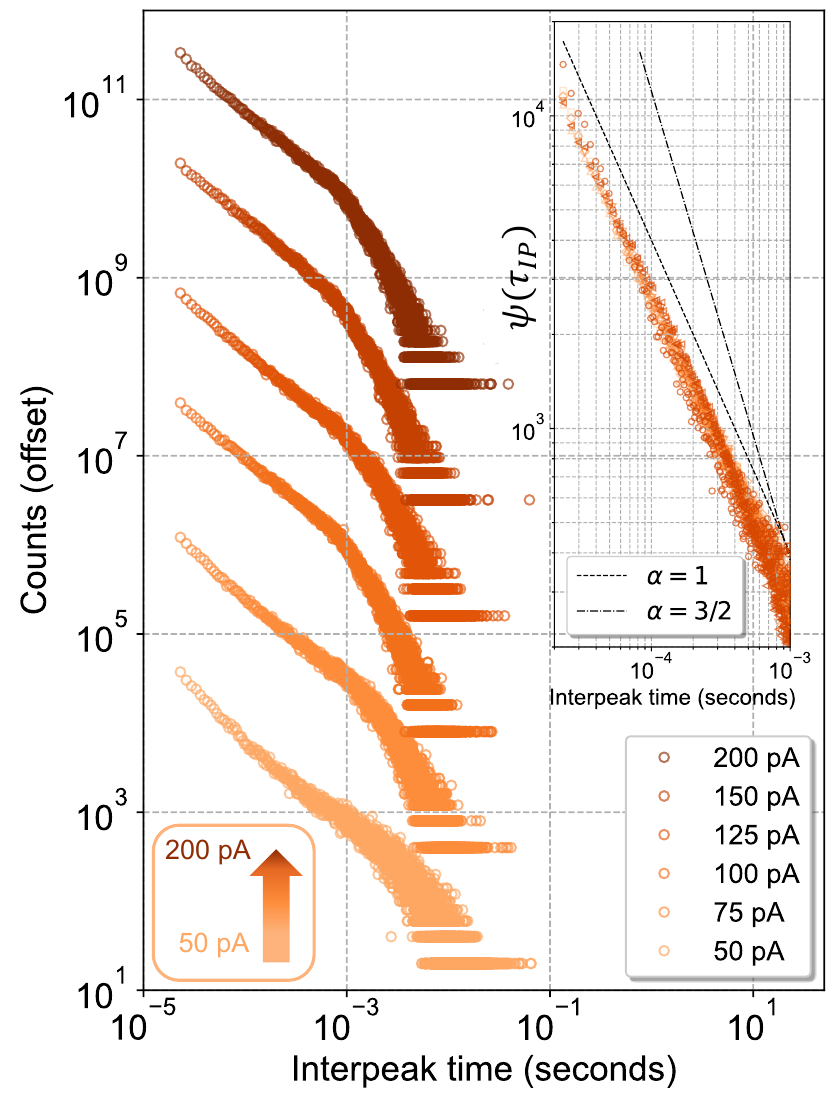}
\caption{\textbf{Measured interpeak time distributions.} Unbinned distributions of interpeak times as a function of setpoint tunnel current, offset along the y-axis for clarity and determined from the same time series as used to generate the residence time distributions in Fig. 2. \textit{Inset:} Zoom of the ITDs for times between 20 $\mu$s (close to the measurement badnwidth limit) and 1 ms, plotted as probability density functions (PDFs) normalised to unity within that time range. No other rescaling (or re-binning) has been used. Dotted and dashed-dotted lines are guides to the eye to highlight gradients of -1 and $-\frac{3}{2}$, respectively, on the log-log plot.}
\label{fig:3}
\end{figure}
\newpage 

\begin{figure}
\centering
\includegraphics[width=1.0\linewidth]{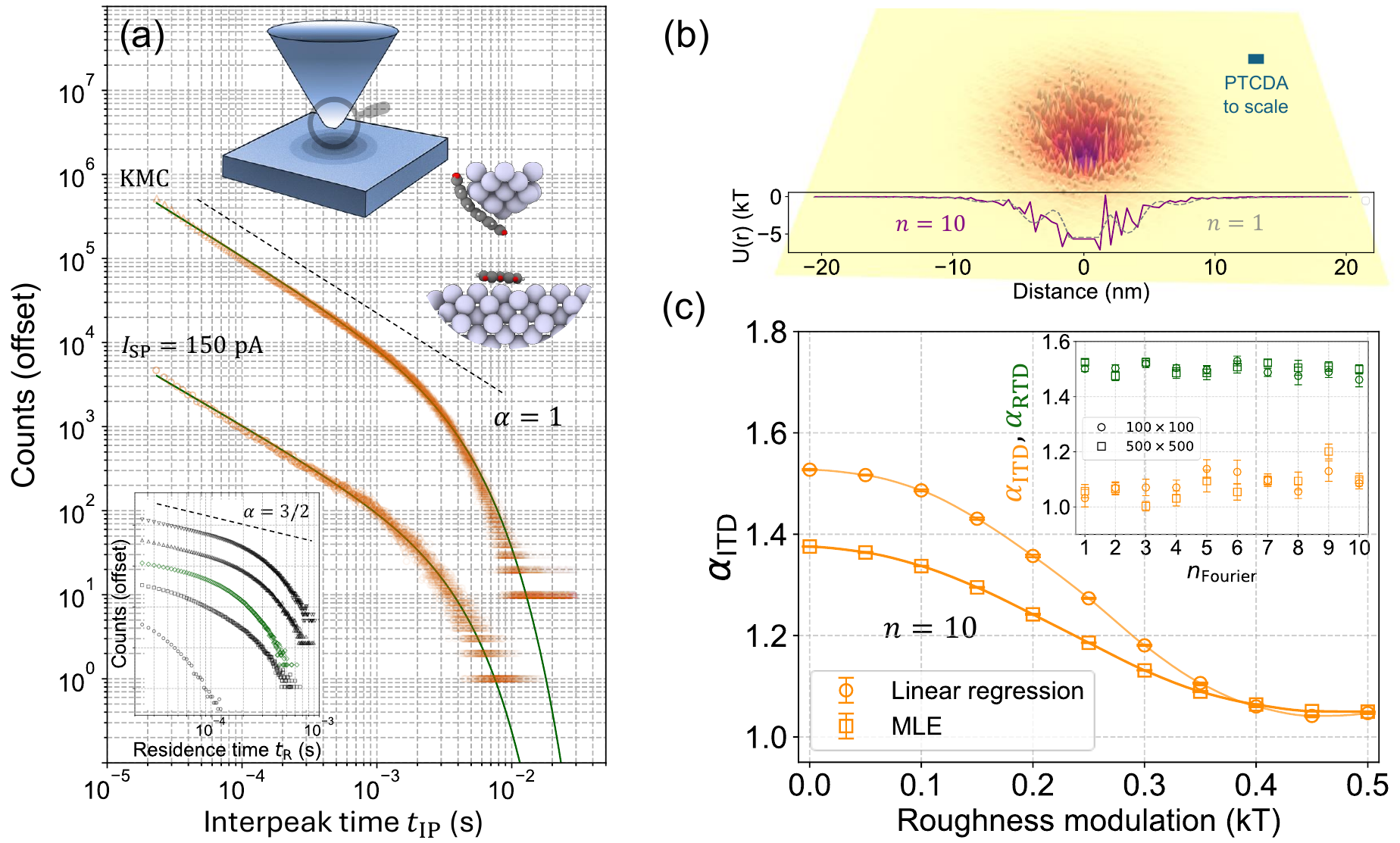}
\caption{\textbf{Emergence of the $\alpha \sim 1$ regime due to heterogeneity in the energy landscape.} \textbf{(a)} Comparison of interpeak time distribution (ITD) produced by KMC simulation (upper curve, open triangles) with experimentally measured ITD for $I_{\mathrm{SP}}=150$~pA (lower curve, open circles). As was the case for Fig. 2 and Fig. 3, the distributions are both plotted in raw, unbinned form. Diffusion coefficients for simulation: $D_{110}=D_{100}=5000$ nm$^2$/s (i.e. isotropic diffusion); 300 kHz sampling was used in the KMC simulation to match the experimental measurement methodology. Both the KMC-generated and measured ITDs have been fit with a KWW-tempered power law (green solid lines) yielding the parameters given in Table 4; \textbf{Inset, upper right:} Schematic of tip–sample geometry; DFT-predicted structure for PTCDA adsorbed on both tip and sample.  \textbf{Inset, lower left:}~KMC-generated~RTDs as a function of simulated detector area. From bottom to top: 0.12,~1.0,~2.6~(green symbols; value used in simulations),~4.0,~5.0~nm$^2$. Dashed lines are each a guide to the eye to highlight the gradient corresponding to $\alpha =1$ or $\alpha=3/2$. \textbf{(b)} Pseudo-three-dimensional image of potential used in the KMC simulation that generated the ITD in (a). A lateral cross section, centred at the tip/dectector apex, is superimposed both for a potential generated from $n=10$~(solid~purple ~line) and $n=1$~(dashed~grey~line) Fourier modes. The blue rectangle shows the relative size of the simulated PTCDA molecule. \textbf{(c)} Variation of $\alpha_{\mathrm{ITD}}$ as determined from \textit{open circles:}~linear~regression, or \textit{filled squares:}~maximum~likelihood ~estimation, as a function of the modulation of the roughness of the potential, $\epsilon$, where \mbox{$U(\mathbf{\rho})=U_0(\mathbf{\rho})(1+\epsilon\phi(\mathbf{\rho}))$} for fixed $\lambda_{\mathrm{c}}=0.5$~nm and $\lambda_{\mathrm{\sigma}}=0.25$~nm. (The form of the roughness field, $\phi(\mathbf{\rho})$, is described in the main text.) In each case analysis was limited to the regime $20~\mu\text{s}\leq t_{\mathrm{IP}}\leq 1\text{~ms}$. The solid lines are spline interpolation provided simply as a guide to the eye and do not imply an underlying model or fit. Error bars (which are smaller than the size of the data markers) are the standard error in the mean for ten KMC runs, each of $10^7$ steps, for a fixed random seed for the roughness field. \textbf{Upper inset:} Insensitivity of $\alpha_{\mathrm{ITD}}$ and $\alpha_{\mathrm{RTD}}$ to the number of Fourier modes for fixed values of $\epsilon=0.4~k_{\mathrm{B}}T$, $\lambda_{\mathrm{c}}=0.5$~nm, and $\lambda_{\mathrm{\sigma}}=0.25$~nm. Error bars are again the standard error in the mean from 10 KMC runs, each of $10^7$ steps, with a different random seed (i.e. separate realisation) for the roughness field each run. We include data for both a $100 \times 100$ and a $500 \times 500$ lattice (open circles or squares, respectively) to highlight that finite size effects do not modify the exponent.}
\label{fig:4}
\end{figure}
\clearpage

\FloatBarrier


\begin{table}
\caption{Results of fitting the RTDs of Fig.~2(a) to an exponentially truncated power law, \mbox{$N(t_{\mathrm{RT}}) \propto t_{\mathrm{RT}}^{-\alpha} \exp(-t_{\mathrm{RT}}/t_{\mathrm{c}})$}.
Here $t_{\mathrm{RT}}$ is the residence time and $t_c$ is a characteristic time scale beyond which long residence times are exponentially suppressed. Reported uncertainties correspond to the 16\%--84\% bootstrap confidence interval, approximately equivalent to a $\pm1\sigma$ range for a Gaussian distribution. Note that $t_c$ increases overall with setpoint, although the 125~pA data point deviates from this trend.}
\centering
\label{tab:1}
\begin{tabular}{ccc}
\hline
Setpoint (pA) & $\alpha$ & $t_c$ (ms) \\
\hline
50  & $1.581\;[1.596,\,1.563]$ & $0.562\;[0.521,\,0.601]$ \\
75  & $1.549\;[1.560,\,1.538]$ & $0.906\;[0.843,\,0.976]$ \\
100 & $1.522\;[1.531,\,1.513]$ & $1.316\;[1.219,\,1.419]$ \\
125 & $1.499\;[1.509,\,1.489]$ & $0.782\;[0.743,\,0.830]$ \\
150 & $1.522\;[1.532,\,1.515]$ & $1.402\;[1.311,\,1.531]$ \\
200 & $1.518\;[1.526,\,1.511]$ & $1.875\;[1.723,\,2.079]$ \\
\hline
\end{tabular}
\vspace{2mm}

\end{table}

\newpage

\begin{table}
\caption{Discrete maximum-likelihood model comparison for the $I_{\mathrm{SP}}=150$ pA interpeak-time distribution ($N = 146{,}909$ events within the fitting window, $20~\mu\text{s}\leq t_{\mathrm{IP}} \leq t_{\mathrm{max}}$, where $t_{\mathrm{max}}$ is set by the largest measured $t_{\mathrm{IP}}$ value (in this case, 78 ms)). 
Models are ranked using differences in the Akaike information criterion ($\Delta$AIC)\cite{Akaike1974} and Bayesian information criterion ($\Delta$BIC)\cite{Schwarz1978} relative to the best-performing model (and reported rounded to the nearest integer). Both criteria penalize model complexity through the number of fitted parameters, $n_{\mathrm{params}}$, discouraging overfitting; lower values indicate stronger empirical support. The $\Delta$AIC and $\Delta$BIC values obtained here provide decisive evidence that the interpeak-time distribution is best described by a KWW-tempered power-law form rather than by conventional alternatives such as Weibull, lognormal, or pure power-law models. See Fig. 4(a) for the corresponding KWW-tempered fit to the data.
The sampling interval is $\Delta t = 3.33\,\mu\mathrm{s}$ with a preamplifier bandwidth of 80 kHz\cite{methods}}
\centering

\begin{tabular}{lccc}
\hline
Model & $n_{\mathrm{params}}$ & $\Delta$AIC & $\Delta$BIC \\
\hline
KWW-tempered power law ($\alpha, t_c, \beta$) 
& 3 & 0 & 0 \\

Exponentially truncated power law ($\alpha, t_c$, $\beta=1$) 
& 2 & 78 & 68 \\

Weibull (KWW)  
& 2 & 2{,}952 & 2{,}942 \\

Mixture of two exponentials 
& 3 & 6{,}431 & 6{,}431 \\

Lognormal ($\mu, \sigma$) 
& 2 & 9{,}684 & 9{,}674 \\

Power law ($\alpha$) 
& 1 & 45{,}826 & 45{,}806 \\

\hline
\end{tabular}

\vspace{2mm}

\label{tab:model_comparison}
\end{table}
\newpage

\begin{table}
\caption{Parameters extracted by maximum likelihood estimation for a KWW-tempered power law description of the measured interpeak time distributions, $N(t_{\mathrm{IP}}) \propto t_{\mathrm{IP}}^{-\alpha_{\mathrm{ITD}}}\exp(-(t_{\mathrm{IP}}/t_{\mathrm{c}})^{\beta}).$ The 16\%-84\% confidence intervals are given. Averaged across all six data sets \mbox{$\alpha_{\mathrm{ITD}}=0.94 \pm 0.04$}, where the error bar is simply $\pm~1\sigma$.}
\centering
\begin{tabular}{lccc}
\hline
 $I_{\mathrm{SP}}$ & $\alpha_{\mathrm{ITD}}$ & $\beta$ & $t_{\mathrm{c}}$ (ms) \\
\hline \\
50~pA  & $0.993^{+0.004}_{-0.002}$ & $2.180^{+0.039}_{-0.021}$ & $15.224^{+0.155}_{-0.145}$ \\\\
75~pA  & $0.974^{+0.002}_{-0.002}$ & $1.622^{+0.022}_{-0.008}$ & $7.566^{+0.057}_{-0.050}$ \\\\
100~pA & $0.979^{+0.002}_{-0.002}$ & $1.427^{+0.017}_{-0.007}$ & $5.264^{+0.037}_{-0.038}$ \\\\
125~pA & $0.918^{+0.002}_{-0.002}$ & $1.221^{+0.011}_{-0.007}$ & $3.576^{+0.020}_{-0.027}$ \\\\
150~pA & $0.933^{+0.002}_{-0.000}$ & $1.170^{+0.011}_{-0.003}$ & $3.002^{+0.018}_{-0.017}$ \\\\
200~pA & $0.871^{+0.002}_{-0.000}$ & $1.018^{+0.009}_{-0.002}$ & $1.965^{+0.009}_{-0.012}$ \\\\
\hline\\
\end{tabular}

\end{table}
\newpage

\begin{table}[b!]
\caption{Comparison of fitted parameters for the KWW-tempered power-law model applied to experimentally measured and KMC-simulated ITDs in the range \mbox{$20~\mu\text{s}\leq t_{\mathrm{IP}}\leq 1~\text{ms}$}. Uncertainties correspond to bootstrap 16\%-84\% confidence intervals. See also table S1 for model comparisons within an information criteria framework. KMC parameters: one molecule, $100 \times 100$ lattice, \mbox{$D_{110}=D_{100}=5000~\text{nm}^2\text{s}^{-1}$}, vdW well depth $6~k_{\mathrm{B}}T$, $\epsilon=0.4~k_{\mathrm{B}}T$, $\lambda_{\mathrm{c}}=~0.5~\text{nm}$, $\lambda_{\mathrm{\sigma}}$=~0.25~\text{nm}, $n_{\mathrm{Fourier}}=10$.}
\centering
\begin{tabular}{lcc}
\hline
\textbf{Parameter} & \textbf{Experiment} & \textbf{KMC simulation} \\
\hline
$\alpha$ & $0.933\,[0.933,\,0.935]$ & $0.995\,[0.992,\,0.996]$ \\
$\beta$ & $1.170\,[1.167,\,1.181]$ & $1.077\,[1.066,\,1.080]$ \\
$t_c$ (ms) & $3.002\,[2.985,\,3.020]$ & $3.194\,[3.156,\,3.202]$ \\
\hline
\end{tabular}
\vspace{2mm}

\label{tab:4}
\end{table}


\clearpage 

%
\bibliography{Moller_et_al} 
\bibliographystyle{sciencemag}

%
%
%
%
%
%

\newpage 
\section*{Acknowledgments}
We thank David Corcoran, Ian Clancy, Brian Kiraly, Philipp Maass, and Alex Saywell for helpful discussions and for their insightful feedback on the initial draft of the manuscript. PM is also grateful to Adam Sweetman for first bringing his attention to Ref.\cite{Shalizi_powerlaw_2007}. 

\paragraph*{Funding:}
PM acknowledges the Engineering and Physical Sciences Research Council (EPSRC) for an Established Career Fellowship (EP/T033568/1). MM is grateful for the award of a Marie Sklodowksa Curie fellowship funded by the ACRITAS FP7 initial training network under the Marie Sklodowska-Curie Actions, project reference ACRITAS-317348-Ares(2016) 909492. PR was funded by a Marie Sklodowska Curie fellowship via the People Programme (Marie Curie Actions) of the European Union’s Seventh Framework Programme (FP7/2007-2013) under Research Executive Agency (REA) Grant No. 628439. The density functional theory (DFT) calculations have been supported by the EPSRC Programme Grant ‘Metal Atoms on Surfaces and Interfaces (MASI) for Sustainable Future’ (EP/V000055/1, EB)  and the University of Nottingham’s Augusta HPC service (SG, EB) and the Sulis Tier 2 HPC platform funded by EPSRC Grant EP/ T022108/1 and the HPC Midlands+ consortium (SG, EB).

\paragraph*{Author contributions:}
The experimental set-up was due to PR; the experiments were carried out by MM and PR. Data analysis was carried out by MM, PR, and PM. Monte Carlo simulations were written by PM (who credits OpenAI's ChatGPT, later replaced by Anthropic's Claude, with a number of coding tasks -- see statement regarding AI usage in \textit{Materials and Methods}.) Density functional theory calculations were carried out by SG, with supervision from EB. The first draft of the manuscript was written by PM and all authors contributed to the final submitted version.

\paragraph*{Competing interests:}
There are no competing interests to declare.

\paragraph*{Data and materials availability:}
Raw data and analysis/simulation codes (written in \textsc{Python} and provided in the form of \textsc{Jupyter} notebooks) are available from \href{10.6084/m9.figshare.31942035}{10.6084/m9.figshare.31942035}

\paragraph{AI usage statement:} OpenAI's ChatGPT (versions GPT-4o and 5-2), later replaced by Anthropic's Claude (Opus 4.6), was used by PM in an iterative process to develop code for: (i)~conversion of Matlab scripts (written by PR) to Python for loading and parsing tunnel current time series recorded in LabVIEW format; (ii)~data plotting and fitting, including maximum likelihood estimation and information criteria metrics; and (iii)~2D kinetic Monte Carlo simulations. These tools were also used to assist with copy-editing and for the conversion of mathematical formulae in the code to raw \LaTeX. All AI-generated outputs were reviewed, verified, and edited by the corresponding author (PM), who takes full responsibility for the content.


\subsection*{Supplementary materials}
Materials and Methods\\
Supplementary Text\\
Figs. S1 to S14\\
Table S1\\
Movie S1\\


\newpage


\renewcommand{\thefigure}{S\arabic{figure}}
\renewcommand{\thetable}{S\arabic{table}}
\renewcommand{\theequation}{S\arabic{equation}}
\renewcommand{\thepage}{S\arabic{page}}
\setcounter{figure}{0}
\setcounter{table}{0}
\setcounter{equation}{0}
\setcounter{page}{1} 


\begin{center}
\section*{Supplementary Materials for\\ \scititle}

M. M\o ller, P. Rahe, S. Ghaderzadeh, E. Besley, and P. Moriarty$^\ast$\\ 
\small$^\ast$Corresponding author. Email: philip.moriarty@nottingham.ac.uk\\
\end{center}

\subsubsection*{This PDF file includes:}
Materials and Methods\\
Supplementary Text\\
Figures S1 to S14\\
Table S1
Caption for Movie S1

\newpage


\subsection*{Materials and Methods}
\subsubsection*{Sample preparation} PTCDA/Ag(110) samples were prepared, and all STM measurements were acquired (at room temperature), in a ScientaOmicron variable temperature ultrahigh vacuum scanning tunnelling microscope-atomic force microscope (VT STM/AFM) system with a base pressure of $2 \ \times 10^{-10}$ mbar. The Ag(110) surface was cleaned using repeated cycles of Ar$^+$ ion sputtering (energy:~800~eV, 10~minutes) followed by annealing at 650~($\pm$~30)~K for 20~minutes. STM images of the Ag(110) surface prepared in this way showed wide ($\sim$ hundreds of nm square), atomically flat terraces with a low defect density. A submonolayer coverage of PTCDA (Sigma-Aldrich) was subsequently deposited via thermal sublimation at 560~($\pm 10$)~K from a home-built source.

\subsubsection*{Tunnel current sampling and atom tracking} The standard current preamplifier of the Scienta Omicron VT STM/AFM instrument was used throughout. For the measurement of tunnel current fluctuations we selected the low gain setting of the preamplifier, as this provides a higher bandwidth: 80 kHz; as such, the time resolution was $\sim$~12~ $\mu$s. A 12 bit analog-to-digital converter (ADC) channel (bandwidth: 300 kHz) of a National Instruments data acquisition (DAQ) card was in turn used to sample the output of the tunnel current preamplifier. Data were acquired over the course of two minutes. In order to maintain the tip position during measurements, an atom tracking unit\cite{Rahe2011} was used to correct for the drift vectors.

\subsubsection*{Generation of RTDs and ITDs} To extract residence time distributions and interpeak time distributions from the fluctuations in the tunnel current time series, we followed the strategy of Ikonomov \textit{et al.}\cite{Ikonomov2010b} and Hahne \textit{et al.}\cite{Hahne2013,HahneThesis}. Briefly, a threshold tunnel current was determined on the basis of a histogram of tunnel current, $I_{\mathrm{T}}$, values whose peak (due to the setpoint tunnel current) was shifted to $I_{\mathrm{T}}=0$. Negative $I_{\mathrm{T}}$ values then arise from noise unrelated to PTCDA diffusion. A Gaussian function, of zero mean, is fitted to the negative values, extending to positive values. We subsequently define the threshold tunnel current, $I_{\mathrm{th}}$ in Fig. 1, on the basis of where the Gaussian distribution count falls below a value of 1 (fig S1; Supplementary Text). We checked that the choice of threshold current (between 0.8~nA and 1.2~nA) did not influence the residence time and interpeak distributions (see Supplementary Text). \\ 

\noindent Having determined $I_{\mathrm{th}}$, we binarise the tunnel current time series to produce a discrete binary time series, $I_\mathrm{B}(t)$, 

\begin{equation*}
I_{\mathrm{B}}(t)= \begin{cases}
1, & \text{if } I_{\mathrm{T}}(t) > I_{\mathrm{th}} \\
0, & \text{otherwise}
\end{cases}
\end{equation*}
\\
\noindent This binarised signal is in turn processed using a simple algorithm that detects when a $0 \rightarrow 1$ or a $1 \rightarrow 0$ change occurs and calculates the RTD and ITD accordingly. Due to the two-minute acquisition period and operation at room temperature, the total number of distinct RTD and ITD values is high, facilitating detailed statistical analyses. Specifically, the total number of residence time values for each RTD, $N_{\mathrm{RT}}$ is as follows:
\begin{center}
\begin{tabular}{r l}
$N_{\mathrm{RT}}$ & $I_{\mathrm{sp}}$ \\
\hline
97,838  & 50~pA \\
163,962 & 75~pA \\
208,771 & 100~pA \\
242,321 & 150~pA \\
280,540 & 175~pA \\
327,007 & 200~pA \\
\end{tabular}
\end{center}
\noindent (Note that $N_{\mathrm{IP}}=N_{\mathrm{RT}}\pm 1$). All data processing/analysis was carried out using \textsc{Python} within a \textsc{Jupyter Notebook} environment. (See \textit{Data and materials availability}.)

\subsubsection*{Binning strategy}  Throughout the main paper we show the experimental data in as ``raw'' a form as possible, i.e. counts~vs~time, with no binning (either linear or logarithmic). In order to provide as close a comparison as possible between the residence/interpeak time distributions generated by the kinetic Monte Carlo (kMC) simulations (see sub-section below) and those obtained in experiment, we also implemented 300~kHz sampling in the kMC code (thus providing counts vs time distributions in an analogus fashion to those measured experimentally). Log-binning has been used only for fig. s4 to accentuate the logarithmic tail in the simulated ITDs that arises from finite size effects/loss of ``memory''.   

\subsubsection*{Maximum likelihood estimation (MLE)} In order to determine the most likely functional form of the experimentally-measured ITDs we used maximum likelihood estimation (MLE)\cite{Clauset2009} for a variety of candidate distributions. MLE was applied directly to the unbinned interpeak times. As the measurements involved sampling with a 300 kHz analog-to-digital converter (i.e. \mbox{$\Delta t=3.33 \mu$s}), the likelihood was evaluated on a discrete-time basis rather than assuming a continuous distribution. For a candidate model with probability density $p(t|\theta)$, where $\theta$ is the parameter set, the probability of observing an interpeak time of $t_k=k\Delta t$ within a fitting window $t_{\mathrm{min}} \leq t_{\mathrm{k}} \leq t_{\mathrm{max}}$ is\\
\begin{equation*}
P_{\mathrm{k}}(\theta)=\frac{p(t_{\mathrm{k}}|\theta)}{\Sigma_{j=k_{\mathrm{min}}}^{j=k_{\mathrm{max}}}p(t_j|\theta)}
\end{equation*}
\\where the denominator provides normalisation. If $N_k$ instances of interpeak time $k\Delta t$ are measured, the log-likelihood is given by
\begin{equation*}
\ln \mathcal{L}(\theta)=\Sigma_{k=k_{\mathrm{min}}}^{k=k_{\mathrm{max}}}N_k \ln P_{\mathrm{k}}(\theta)
\end{equation*}
In practice, the likelihood was calculated using the unique counts and interpeak time values measured experimentally. Parameter estimates were obtained by maximising $\ln \mathcal{L}(\theta)$. Fixed values of $t_{\mathrm{min}}=20 \mu$s and $t_{\mathrm{max}}=1$ ms were used throughout (matching the window shown in the inset~to~Fig.~3).\\

\noindent We subsequently compared multiple candidate models using the Akaike\cite{Akaike1974} and Bayesian\cite{Wit2012,Schwarz1978} information criteria, AIC and BIC, respectively, defined as follows:
\begin{equation*}
\mathrm{AIC}=2n_{\mathrm{params}} - 2 \ln L_{\mathrm{max}} \qquad \mathrm{BIC} = n_{\mathrm{params}} \ln N - 2 \ln L_{\mathrm{max}}
\end{equation*}

\noindent where $n_{\mathrm{params}}$ is the number of model parameters, $L_{\mathrm{max}}$ is the maximised likelihood, and $N=\Sigma_{k=k_{\mathrm{min}}}^{k=k_{\mathrm{max}}}N_k $ is the total number of interpeak time values within the fitting window. Differences in AIC and BIC were used to evaluate support for a given model, with smaller values in each case indicating better agreement between the empirical data and the model. AIC and BIC yielded consistent rankings (Table 2). 
\subsubsection*{Density functional theory} The density functional theory (DFT) calculations were performed with the Vienna Ab initio Simulation Package (\textsc{VASP})\cite{Kresse1996,Kresse1999}, within the plane-wave projector augmented-wave (PAW) method. The structures were relaxed using the Perdew–Burke–Ernzerhof (PBE) exchange-correlation functional\cite{Perdew1996} with a force tolerance of 0.01 eV \AA$^{-1}$ and an electronic convergence criteria of $10^{-6}$ eV. The energy cut-off was set to 500 eV. Given that the simulation supercell was large – containing nearly 400 metal atoms along with other elements – a Monkhorst-Pack k-point grid of 1 × 1 × 1 was used to sample the Brillouin zone. Van der Waals interactions were taken into account using the DFT-D2 method of Grimme\cite{Grimme2006}. The supercell contained 393 silver, 48 carbon, 16 hydrogen, and 12 oxygen atoms, with a total of 4603 valence electrons.

\subsubsection*{Kinetic Monte Carlo simulations} A kinetic Monte Carlo (KMC) simulation was developed (in Python) to model PTCDA diffusion and generate simulated RTDs and ITDs for comparison with experimental data. (A Jupyter Notebook (and associated Python modules) containing the code is available. See \textit{Data and materials availability}.) Our KMC code was inspired by that implemented by Hahne~\textit{et~al.}\cite{HahneThesis,Hahne2014} for numerical modelling of RTDs and ITDs stemming from molecular diffusion but differs from their approach in a number of ways. We incorporated (i) a potential due to the tip (see following section), (ii) a detector geometry that spanned more than one lattice cell, and (iii) intermolecular interactions via a simple empirical Morse potential. (Ultimately, the inclusion of intermolecular forces was not necessary to accurately reproduce the experimental ITDs. (Supplementary Text.) Moreover, our KMC strategy models the 300 kHz sampling used in experiment so as to produce raw counts vs $t_{\mathrm{IP}}$ (or $t_{\mathrm{RT}}$) distributions for direct comparison with their measured counterparts.\\

\noindent Molecular diffusion was simulated using a rejection-free kinetic Monte Carlo (KMC) algorithm on a discrete lattice with periodic boundary conditions. For a molecule located at lattice site $(i,j)$, transitions to the four nearest-neighbour sites $(i\pm1,j)$ and $(i,j\pm1)$ are allowed. Each possible hop $k$ is assigned a direction-dependent rate of the form
\begin{equation*}
\nu_k = \gamma_\mu \exp\!\left(-\frac{\Delta U_k}{2k_{\mathrm B}T}\right),
\end{equation*}
where $\mu \in \{x,y\}$ denotes the hop direction, $\gamma_\mu = D_\mu/a_\mu^2$ is the directional attempt frequency defined by the diffusion coefficient $D_\mu$ and lattice spacing $a_\mu$, and $\Delta U_k = U_{i+\delta_i,\,j+\delta_j} - U_{i,j}$ is the change in the tip-induced potential associated with that move.\\

\noindent This symmetric (local detailed-balance) form ensures that the ratio of forward and reverse transition rates satisfies the Boltzmann condition, so that the correct equilibrium distribution is recovered. In particular, even when the lattice discretization is anisotropic ($a_x \neq a_y$), isotropic diffusion in physical space is obtained by scaling the directional rates according to $\gamma_\mu = D_\mu/a_\mu^2$.\\

\noindent The total escape rate $\nu_{\mathrm{tot}} = \sum_k \nu_k$ is used to generate the time increment
\begin{equation*}
\Delta t = -\frac{\ln u}{\nu_{\mathrm{tot}}},
\end{equation*}
where $u \in (0,1)$ is a uniform random number. The direction of the hop is then selected with probability $\nu_k / \nu_{\mathrm{tot}}$, yielding a continuous-time, rejection-free trajectory consistent with the underlying potential landscape.

\subsubsection*{Choice of base diffusion coefficient} Our choice of diffusion coefficient in the KMC simulations was initially informed by the mean residence times observed in experiment. A value of $D=5000~\text{nm}^2/\text{s}$ is consistent with the experimentally measured  mean residence times, \mbox{$\langle t_{\mathrm{RT}}\rangle \sim 50 - 70 ~\mu\text{s}$}, since this corresponds to an rms diffusion length of order 1 nm, comparable to the molecular (and detector) dimensions (see below). It is also in line with the value of \mbox{$D\sim1.25~\times~10^4$ nm sqr/s} determined by the analysis of Hahne \textit{et al.}\cite{Hahne2013} for PTCDA diffusion on Ag(100); the somewhat lower diffusion constant here is reasonable given the lower atomic packing density of the Ag(110) surface. Although, as described above, the KMC simulations can model anisotropic hopping, i.e. $D_{[110]}\neq D_{[100]}$, we found, somewhat unexpectedly, that diffusion anisotropy did not improve agreement with experiment (Supplementary Text; fig S9). Instead, the directional bias in molecular trajectories effectively reduced the spatial homogeneity of the dynamics within the detection region and degraded the agreement with the observed RTD and ITD scaling. For this reason, isotropic diffusion was used throughout. (Moreover, distinguishing between isotropic and anisotropic diffusion on the basis of RTDs and ITDs alone is exceptionally problematic (at best) unless the STM tip is dithered in an oscillatory trajectory\cite{HahneThesis}.)

\subsubsection*{KMC Detector Geometry} As described in the body of the paper (see, in particular, the discussion related to the lower inset in Fig. 4(a)), use of a single lattice site as a detector\cite{HahneThesis, Hahne2014} did not yield the $\alpha \sim 3/2$ scaling regime clearly observed in the measured RTDs. We therefore implemented a detection geometry that provided a more diffuse boundary and also took account of the convolution of the tip and molecule ``footprints''. The detector was represented by a weighting field centred at the tip apex rather than by a single lattice site. Specifically, a Gaussian sensitivity function
\begin{equation*}
w(\mathbf{r})=
\exp\!\left[
-\left(
\frac{x^2+y^2}{2\sigma^2}
\right)
\right]
\end{equation*}
was defined on the periodic simulation lattice, where $\mathbf{r}=(x,y)$ is the position relative to the detector centre and $\sigma$ represents the detector width. For a molecule centred at lattice site $\mathbf{R}_m$, with footprint offsets $\{\boldsymbol{\delta}_k\}$ describing the occupied lattice cells of the molecule, the detector ``score'' was
\begin{equation*}
S(\mathbf{R}_m)=\sum_k w(\mathbf{R}_m+\boldsymbol{\delta}_k).
\end{equation*}
The detector output was then binarised according to
\begin{equation*}
\chi(\mathbf{R}_m)=
\begin{cases}
1, & S(\mathbf{R}_m)\ge S_{\mathrm{thr}},\\[4pt]
0, & S(\mathbf{R}_m)< S_{\mathrm{thr}},
\end{cases}
\end{equation*}
where $S_{\mathrm{thr}}$ is the detection threshold.\\

\noindent To characterise the detector size, we evaluated the score field for a molecule centred at every lattice site,
\begin{equation*}
S_{\mathrm{center}}(\mathbf{r})=\sum_k w(\mathbf{r}+\boldsymbol{\delta}_k),
\end{equation*}
and defined the effective detection area from the region for which $S_{\mathrm{center}}(\mathbf{r})\gtrsim S_{\mathrm{thr}}$. 


\subsubsection*{Tip potential}
The tip–molecule interaction was modelled using a nonretarded sphere–plane van der Waals form, motivated by the integrated tip–adsorbate geometry discussed by Olsen \textit{et al.}\cite{Olsen2012} (see also Israelachvili\cite{Israelachvili2010}) for an adsorbate interacting with a curved STM tip above a planar surface. Writing $\rho$ for the lateral distance from the tip apex, the potential was taken to be
\begin{equation*}
U_{\mathrm{vdW}}(\rho)
=
-\frac{C}{d(\rho)^3},
\qquad
d(\rho)=\sqrt{\rho^2+(R_{\mathrm{tip}}+h)^2}-R_{\mathrm{tip}},
\end{equation*}
where $R_{\mathrm{tip}}$ is the tip radius of curvature (set at $30~\text{nm}$ for the simulations described in the main text), and $h$ is the tip--surface gap (which fell in the range 0.5 to 1.0 nm.) (As Olsen \textit{et al.}\cite{Olsen2012} highlight, this is an upper limit.) The prefactor, $C$, was chosen so that the on-axis potential depth at a reference gap $h_{\mathrm{ref}}=0.7~\text{nm}~$was equal to $-U_{\mathrm{ref}}$, giving
\begin{equation*}
C = U_{\mathrm{ref}}\,h_{\mathrm{ref}}^{3},
\end{equation*}
and hence
\begin{equation*}
U_{\mathrm{vdW}}(\rho)
=
-\,U_{\mathrm{ref}}
\left(
\frac{h_{\mathrm{ref}}}
{\sqrt{\rho^2+(R_{\mathrm{tip}}+h)^2}-R_{\mathrm{tip}}}
\right)^3.
\end{equation*}
In particular, on axis this reduces to
\begin{equation*}
U_{\mathrm{vdW}}(0)
=
-\,U_{\mathrm{ref}}\left(\frac{h_{\mathrm{ref}}}{h}\right)^3.
\end{equation*}
A flat-core region with smooth blending to the asymptotic sphere--plane form was introduced near the tip centre in order to preserve the short-time scaling of the RTDs (as discussed in the main text). Spatial roughness was then incorporated as a multiplicative modulation of this baseline potential. (Again, see main text).
\newpage


\subsection*{Supplementary Text}

\subsubsection*{Choice of threshold tunnel current for binarisation}
As noted in \cite{methods}, the tunnel current was sampled from the I-V converter of the Scienta Omicron Variable Temperature STM (VT-STM) using a 12-bit National Instruments (NI) analog-to-digital converter (ADC). The I-V converter/preamplifier had a gain of 33.3~nA/V, a bandwidth of 80~kHz, and was sampled by the NI ADC at 300~kHz. The 12-bit ADC combined with the I-V converter gain results in a tunnel current quantisation step of (20~V/2$^{12}$)(33.3~nA/V), i.e. $\sim$ 163 pA.  We use the same strategy as that adopted by Hahne \textit{et al.}\cite{Hahne2013, HahneThesis} in order to determine the value of $I_{\mathrm{th}}$ (Fig. S1). Negative current values are associated with instrumental noise and are distinct from the fluctuations due to molecular diffusion. A Gaussian is fit to the low tunnel current values, as shown in Fig. S1, to account for the instrumental noise that is distinct from molecular diffusion. (See the following section, \textit{Influence of tip state}, for strong supporting evidence of the efficacy of this strategy.) The threshold, $I_{\mathrm{th}}$, is chosen as the (positive) tunnel current for which the value of the Gaussian is one event.\\
 
\noindent As it is the temporal characteristics of the signal above the binarisation threshold that are of central interest in this study, we have traded tunnel current resolution for bandwidth. However, in order to test that the relatively coarse tunnel current quantisation of $\sim$ 160 pA did not affect the measured residence time distributions (RTDs) and/or interpeak time distributions (ITDs), we have varied $I_{\mathrm{th}}$ by $20\%$ either side of the value determined via the heuristic described in the preceding paragraph. As shown in Fig. S2, the form of both the RTD and ITD is robust against these substantial changes in $I_{\mathrm{th}}$. (In this context, we also highlight Jani\'cevi\'c \textit{et al.}'s\cite{Janicevic2016} important analysis of the spurious correlations that can arise from an inadequate choice of finite detection thresholds.)   

\subsubsection*{Influence of tip state}
Unsurprisingly, the tip state can have a significant influence on the nature of the tunnel current fluctuations (Fig.~S3) and, as such, we ensured that RTDs and ITDs were only extracted from time series of tunnel current fluctuations measured with an appropriate probe state. Note how only Fig.~S3(a) exhibits ``speckle'' noise in the neighbourhood of the PTCDA island; the other tips produce images of Ag(110) terraces that are much cleaner and well-behaved (Figs.~S3(b),~(c)); this is seen more clearly in the corresponding line profiles of Fig.~S3(d). Inset to each of the images in Figs.~S3(a)-(c) is a rudimentary schematic illustration of our interpretation of the tip state; only for (a)~are molecules free to diffuse. Tip~state~c is particularly valuable in the context of a control experiment: no fluctuations other than those arising due to instrumental noise are possible because the probe cannot detect diffusing molecules. As such, the tunnel current histogram (Fig.~S3(e)) comprises only the Gaussian distribution described in the previous section. In addition to this control measurement, we also routinely measured on and off PTCDA islands. There are no diffusing molecules on top of the islands and so we again only record a Gaussian distribution for the tunnel current fluctuations.      

\subsubsection*{Relationship of $P_{\mathrm{occ}}(\ln(I_{\mathrm{SP}}/I_0))$ to $\langle \tau_{\mathrm{IP}}\rangle(\ln(I_{\mathrm{SP}}/I_0))$}

In Fig.~2(b) of the main paper we have fitted a function of the following form to the measured $\langle \tau_{\mathrm{IP}}\rangle(\ln(I_{\mathrm{SP}}/I_0)) $ data:
\begin{equation}
    \langle \tau_{\mathrm{IP}} \rangle(d')=\frac{\langle \tau_{\mathrm{RT}} \rangle}{\alpha + \beta d'}-\langle \tau_{\mathrm{RT}} \rangle 
\end{equation}
where $d'=\ln(I_{\mathrm{SP}}/I_0)$, with $I_{\mathrm{SP}}$ the setpoint current and $I_0=1$~pA (the latter only to formally ensure a dimensionless argument for the log function), which scales linearly with the tip-sample separation. The inverse-linear function of Eqn. S1 stems from the following simple considerations.\\

\noindent First, 
\begin{equation}
P_{\mathrm{occ}}=\frac{\Sigma \tau_{\mathrm{RT}}}{\Sigma \tau_{\mathrm{RT}}+\Sigma \tau_{\mathrm{IP}}}
\end{equation}
But $\Sigma \tau_{\mathrm{RT}}=n_{\mathrm{RT}}\langle \tau_{\mathrm{RT}}\rangle$ and $\Sigma \tau_{\mathrm{IP}}=n_{\mathrm{IP}}\langle \tau_{\mathrm{IP}}\rangle$. Moreover $n_{\mathrm{RT}} \sim n_{\mathrm{IP}} \equiv n$. (Because $n_{\mathrm{RT}}=n_{\mathrm{IP}} \pm 1$, and $n_{\mathrm{IP}}$ is of order 10$^5$.) Hence,
\begin{equation}
    P_{\mathrm{occ}}=\frac{\langle \tau_{\mathrm{RT}} \rangle}{\langle \tau_{\mathrm{RT}} \rangle + \langle \tau_{\mathrm{1P}} \rangle} \implies \langle \tau_{\mathrm{IP}} \rangle = \langle \tau_{\mathrm{RT}} \rangle \left(\frac{1-P_{\mathrm{occ}}}{P_{\mathrm{occ}}}\right)
\end{equation}
The occupation probability, $P_{\mathrm{occ}}$, depends linearly on $d'$ (Fig. 2(c)), i.e. $P_{\mathrm{occ}}(d') =\alpha + \beta d'$, with $\alpha$ and $\beta$ as constants. Hence, from [S3],
\begin{equation}
\langle \tau_{\mathrm{IP}}\rangle(d')=\langle \tau_{\mathrm{RT}}\rangle \left(\frac{1-P_{\mathrm{occ}}(d')}{P_{\mathrm{occ}}(d')}\right)
\end{equation}
or, equivalently,
\begin{equation}
\langle \tau_{\mathrm{IP}}\rangle(d')=\frac{\langle \tau_{\mathrm{RT}}\rangle}{\alpha+\beta d'} - \langle \tau_{\mathrm{RT}}\rangle
\end{equation}

\noindent [S5] is the fitting function specified in [S1] above. We take the mean of $\langle \tau_{\mathrm{RT}}\rangle$ across all six values of $d'$ and use this as a fixed parameter in the fit shown in Fig. 2(c).

\subsubsection*{Maximum likelihood estimation and Bayesian/Akaike information criteria} Table~S1 shows comparisons of six different model distributions for each of the ITDs acquired in experiment (as a function of $I_{\mathrm{SP}}$, i.e. increasing tip-sample interaction.) For all but the $I_{\mathrm{SP}}=200~\text{pA}$ data, the KWW-tempered power law is decisively favoured by the Akaike and Bayesian information criteria metrics. For the $I_{\mathrm{SP}}=200~\text{pA}$ distribution, an exponentially-truncated power law with $\beta=1$, rather than a KWW-tempered variant (i.e. where $\beta \neq 1$), is favoured by the BIC measure.  

\subsubsection*{Comparison of analytical and KMC-generated ITDs for a single, unbiased walker}

Fig. S4 is an interpeak-time probability density function generated from the simulation of a single PTCDA molecule (of size 3~$\times$~5 lattice sites) diffusing isotropically on a 1000~$\times$~1000 lattice in the absence of any external potential or bias. In this case, and as opposed to all other ITDs plotted in the main paper and in this supplement, we have used logarithmic binning (in order to accentuate the exponential tail of the distribution.) The solid line is a fit to the analytical form of the interpeak time probability density function that was derived in \cite{Hahne2013,HahneThesis}:

\begin{equation}
\psi(t_{\mathrm{IP}})=\frac{2}{\pi\tau_{\mathrm{R}}} \int_0^{\infty} d\chi \ \chi \ \frac{W_0[\chi(1+\epsilon), \chi]}{J_0^2(\chi)+Y_0^2(\chi)} \ \exp(-\chi^2t_{\mathrm{IP}}/\tau_{\mathrm{R}})
\end{equation} \vspace{3mm}

\noindent where $\tau_R$ is the mean residence time, $W_0(x,y)=J_0(x)Y_0(y)-J_0(y)Y_0(x)$ with $J_0$ and $W_0$ being zeroth-order Bessel functions of the first and second kind, respectively, $\epsilon$ is a measure of the initial displacement of the diffusing particle from an absorbing boundary, and $\chi$ represents a continuous spectrum of eigenvalues. (See Appendix to \cite{Hahne2013} for a full derivation. (Here we use the form for circular molecules, not least because a closed form for rectangular molecules has not been found to date\cite{Hahne2014,HahneThesis}. For aspect ratios comparable to that of the simulated PTCDA molecule (namely,~1.67), the ITD for the circular molecule case is a very good approximation.) As is clear from Fig.~S4, Eqn.~S6 provides a very close fit to the KMC-generated ITD across five orders of magnitude, up to $\tau_{\mathrm{IP}} \sim$~1~s. 

\subsubsection*{The $\alpha \sim \frac{3}{2}$ regime} In the broader context of bursty dynamics, the full functional form of the fit to the KMC-generated data of Fig.~S4 is arguably of less interest than the limiting behaviour on different time-scales. In particular, and as highlighted by the dotted line in Fig. S4, the limiting form of Eqn.~S6 over a relatively large range of $\tau_{\mathrm{IP}}$ is a truncated power law\cite{Hahne2013,HahneThesis},

\begin{equation}
\psi(t_{\mathrm{IP}})\propto \exp\left(-\frac{\epsilon^2\tau_{\mathrm{R}}}{4t_{\mathrm{IP}}}\right)\left(\frac{t_{\mathrm{IP}}}{\tau_{\mathrm{R}}}\right)^{-3/2} 
\end{equation}

\noindent \\ The power law dependence arises from a first-passage-time (or return time) analysis\cite{Redner2012} in the context of random walk theory. This is especially important in the context of Barab\'asi \textit{et al.}'s queuing models of bursty human dynamics\cite{Barabasi2005, Vazquez2005, Vazquez2006} -- particularly with regard to traditional ``snail mail'' communication (in particular, the correspondence of Einstein, Darwin, and Freud)\cite{Oliveira2005, Vazquez2006} -- because their $\alpha=\frac{3}{2}$ ``universality class'' has essentially the same fundamental origin: a first-passage time problem for a random walker, albeit in 1D, rather than 2D.\\ 

\noindent The $\tfrac{3}{2}$ exponent arises naturally from first-return statistics for an unbiased random walk. After leaving the detection region (which for the simulations that generated Fig. S4 was a single lattice site), the interpeak time is defined as the \emph{first-return time} to that site. The relevant quantity is therefore the survival probability $S(t_{\mathrm{IP}})$, defined as the probability that the walker has not yet returned by time $t_{\mathrm{IP}}$. In this regime (see also ``\textit{Marginal recurrence and finite size effects}'' below), the survival probability follows an approximate scaling $S(t_{\mathrm{IP}})\sim t_{\mathrm{IP}}^{-1/2}$. The interpeak time probability distribution $\psi(t_{\mathrm{IP}})$, being the first-return time density, is given by
\begin{equation}
\psi(t_{\mathrm{IP}}) = -\frac{\mathrm{d}}{\mathrm{d}t_{\mathrm{IP}}} S(t_{\mathrm{IP}}),
\end{equation}
and hence scales as
\begin{equation}
\psi(t_{\mathrm{IP}})\sim t_{\mathrm{IP}}^{-3/2}.
\end{equation}

\noindent This $\alpha \sim \tfrac{3}{2}$ behaviour is a generic signature of diffusion-controlled first-return processes~\cite{Redner2012}, and has been discussed in detail in the context of STM measurements by Hahne \textit{et al.}~\cite{Hahne2013,HahneThesis}.

\subsubsection*{Marginal recurrence and finite size effects} Eqn.~S6 is a good fit to the KMC-generated ITD up until $t_{\mathrm{IP}} \sim 1$ s. Beyond this point, the asymptotic limit of
\begin{equation}
\psi(t_{\mathrm{IP}})\propto\frac{1}{t_{\mathrm{IP}}\ln^2(t_{\mathrm{IP}})}
\end{equation}

\noindent \\ for a 2D random walker on an infinite lattice -- the marginal recurrence limit \cite{Redner2012} -- is superseded by an exponential tail (dashed line in Fig.~S4). Hahne \textit{et al.}\cite{Hahne2013} discuss how the asymptotic behaviour predicted by Eqn.~S10 for a single walker is modified in the presence of additional molecules: a different particle can reach the detection area before the original has time to return from its random walk, giving rise to an exponential decay, rather than the marginal recurrence of Eqn.~4, 
\begin{equation}
\psi(t_{\mathrm{IP}}) \propto c D \exp(-\kappa \pi cD t_{\mathrm{IP}}), 
\end{equation} where $c$ is the molecular concentration, $D$ is the diffusion constant, and $\kappa$ is a constant of order unity. \\  

\noindent However, the simulated ITD of Fig.~S4 has been generated by KMC modelling of a single diffusing molecule, and yet the long-time behaviour is of exponential, rather than marginally recurrent, character. Given that only one molecule is present in the simulation, the exponential tail in Fig. S4 cannot be attributed to contributions from another walker reaching the monitor site first. Instead, this is a finite-size effect arising from the lattice dimensions in the simulation. Periodic boundary conditions result in a diffusion space that is clearly not infinite in extent but is best described as having a toroidal topology. In the large $t_{\mathrm{IP}}$ regime, the mean square displacement of the molecule from the monitor site saturates at a characteristic time set by the size of the simulated lattice -- the random walk has reached the ergodic limit. (See \textit{Mean square displacement} section below). Paralleling the case of multiple molecules, the finite lattice size erases the particle's memory of its trajectory and an exponential tail in the ITD results. Accordingly, the onset of the exponential tail (highlighted with an arrow in Fig.~S4) shifts to longer interpeak times with increasing lattice size, and scales, for a fixed value of $D$, with $L^2$, where $L$ is the lattice side length. \\

\noindent We highlight this difference between the marginal recurrence limit and an exponential decay because it both informs the design of the KMC simulations (in terms of choice of lattice size for a given value of $D$), and, more importantly, Eqn.~S10 in principle provides a mechanism for the observation of $\alpha \sim 1$ scaling. (Eqn.~S10 holds for a 2D random walk. In 1D, the asymptotic limit is $\psi(t_{\mathrm{IP}}) \propto t_{\mathrm{IP}}^{-3/2}$. See Section 1.5 of Ref.\cite{Redner2012}.) However, this would require not only an exceptionally dilute molecular coverage in the STM detection experiment but also wide, defect free Ag(110) terraces: molecular diffusion must be entirely unconstrained so as to reach the marginal recurrence limit (and it is clear from the measured ITDs (Fig. 3 in the main paper) that the tip potential has a significant influence on PTCDA trajectories). More simply, Fig.~3 of the main paper shows that the ``$\alpha_{\mathrm{ITD}} \sim 1$'' regime in experiment occurs in the wrong range of $t_{\mathrm{IP}}$ for this to arise due to marginal recurrence, i.e. at times far below the $t_{\mathrm{IP}} \rightarrow \infty$ limit. 

\subsubsection*{Additional density functional theory (DFT) calculations}
Details of the DFT calculations are given in \cite{methods}. Here we include additional results/relaxed geometries. In each case, the DFT framework was the Vienna Ab Initio Simulation Package (VASP) within the Plane-wave Projector Augmented Wave (PAW) method, supplemented by the DFT-D2 method of Grimme\cite{Grimme2006} to account for dispersion forces. In each case the uppermost two layers of the tip were held in place during relaxation. Fig.~S5 shows the converged geometries for a clean Ag tip above a Ag(110) surface (Fig. ~S5(a)) and above a PTCDA molecule adsorbed on Ag(110) (Fig.~S5(b)). We note that the presence of the molecule reduces the interaction strength as compared to a Ag tip-Ag(110) interaction. \\

\noindent A number of adsorption geometries for PTCDA on the tip were explored, including the ``pointed'' termination shown in Fig. S6. As shown in Fig. S7, with this geometry there is a significant variation of tip-sample binding energy across the PTCDA molecule adsorbed on Ag(110). Fig. S8 shows another tip-adsorbed PTCDA geometry, in this case a ``quasi-planar'' termination given rise to a significantly enhanced interaction strength ($\sim 64~k_{\mathrm{B}}T$ at room temperature. A potential well of that depth would, however, entirely immobilise any surface-adsorbed PTCDA molecule that diffused within sufficient range of the tip apex. As such, we ruled out this geometry for the KMC simulations. It is likely, however, that a geometry of this type is responsible for tip state C shown in Fig. S3. We also include Movie S1 for the ``quasi-planar'' geometry, showing the movement of the surface-adsorbed PTCDA towards the tip.

\subsubsection*{Anisotropic diffusion} Before introducing the roughened potential energy landscape described in the main paper (Fig.~4 and associated discussion), we explored the effects of increasingly anisotropic diffusion on the scaling exponent $\alpha_{\mathrm{ITD}}$ in the presence of a simulated van der Waals well (of a depth of $6 ~k_{\mathrm{B}}T$, $R=30~\text{nm}$, $h=0.7~\text{nm}$) but with the roughness field, $\phi(\rho)$, set to 0. The results are shown in Fig.~S9 for both a $100 \times 100$ and a $500 \times 500$ lattice. Holding the diffusion coefficient along the 110 direction, $D_{110}$, constant at $5000 ~\text{nm}^2\text{s}^{-1}$, we consecutively set the diffusion coefficient in the orthogonal direction, $D_{100}$, to a value of $50, 100, 500, 1000, 5000$, and $10000$ nm$^2$s$^{-1}$ and determined the value of $\alpha_{\mathrm{ITD}}$ in the range 20 $\mu$s to 1 ms via maximum likelihood estimation (across ten KMC runs). It is clear from Fig. S9 that anisotropic diffusion is insufficient to yield $\alpha_{\mathrm{ITD}} \sim 1$ as observed in experiment. 

\subsubsection*{Beyond a single walker}Holding $D_{110}=D_{100}=5000~\text{nm}^2s^{-1}$, with a van der Waals potential well of the same form as described for Note \#7A above, we varied the number of molecules in the KMC simulation from $n=1$ to $n=20$. As seen in Fig.~S10, the value of $\alpha_{\mathrm{ITD}}$ diverges from, rather than approaches, $\alpha_{\mathrm{ITD}} \sim 1$ seen in experiment. The ITDs shown in the lower inset of Fig.~S10 demonstrate why this is the case -- with increasing molecular number density, the exponential tail in the interpeak time distribution is pushed to ever shorter times, artificially increasing the ``slope'' of the distribution in the 20 $\mu$s to 1 ms range of $t_{\mathrm{IP}}$. Although it is of course possible to rescale the time axis to account for this, the net result is the recovery of a value of $\alpha_{\mathrm{ITD}}$ close to 3/2. In other words, and similar to the results of the tests of anisotropic diffusion, we find that increasing molecular number density in the simulations fails to yield a value of $\alpha_{\mathrm{ITD}}$ close to 1.\\ 

\noindent For Fig.~S10, molecular interactions were of a simple ``hard-sphere'' type -- molecules could not overlap, but there was no additional attractive or repulsive interaction. However, the introduction of an intermolecular Morse potential also did not recover the $\alpha \sim 1$ scaling seen in experiment. Intermolecular interactions also led to residence times that increased significantly with decreasing simulated tip-sample separation, again in contradiction to experiment.

\subsubsection*{Variation of $\lambda_{\mathrm{c}}$}
In Fig.~4 of the main paper we show how $\alpha_{\mathrm{ITD}}$ varies both with $\epsilon$, the modulation strength for the roughness field, and with $n_{\mathrm{Fourier}}$, the number of Fourier modes (with all other roughness parameters fixed.) Figs.~S11~and~S12 supplement the data in the main paper by demonstrating the dependence of $\alpha_{\mathrm{ITD}}$ on the target wavelength for the roughness field, with all other parameters (other than the random seed for the choice of roughness parameters) fixed (as listed in the captions to Fig.~S11 and S12). The error bars are standard errors in the mean for ten different KMC simulations (each with a different roughness realisation due only to the random seed.)  \noindent 

\noindent When the wavelengths that characterise the spatial modulation of the roughness field are comparable to both the size of the diffusive hop length (i.e. $a_{110}=0.29~\text{nm},a_{100}=0.4~\text{nm}$) and the scale of the detection region ($\sim 1~\text{nm}$), the run-to-run variability in the value of $\alpha$ is relatively low (Fig.~S11). Conversely, when $\lambda_{\mathrm{c}}$ significantly exceeds the scale of the detection region, both the mean value of $\alpha$ and the scatter increase significantly (Fig.~S12). We interpret this as the first-passage dynamics becoming less ``self-averaging'' as the characteristic modulation wavelength grows -- each realisation of the roughness field yields a very different local environment for large $\lambda_{\mathrm{c}}$.   

\subsubsection*{Mean square displacement}
It is well established that diffusion on heterogeneous energy landscapes can deviate strongly from ideal Brownian behaviour, exhibiting anomalous scaling of the mean square displacement (MSD), $\langle r^2(t)\rangle \propto t^{\gamma}$ with $\gamma \neq 1$\cite{BouchaudGeorges1990,Bouchaud2012}. Similarly, stretched (or compressed) exponentials typically reflect an underlying heterogeneous/disordered energy landscape and are often accompanied by non-Brownian behaviour\cite{Phillips1996,Laherrere1998,Chen2003,Wu2016}. As such, we computed the MSD over the $20~\mu\text{s} \leq t_{\mathrm{IP}} \leq 1~\text{ms}$ window for a single simulated PTCDA molecule under three different conditions: (i) free diffusion (i.e. no tip potential), (ii) a smooth van der Waals well of depth $6~k_{\mathrm{B}}T$ (see upper inset to Fig.~S10), and (iii) a vdW well of the same depth modulated by a roughness field defined by following parameters: $\epsilon=0.4~k_{\mathrm{B}}T$, $\lambda_{\mathrm{c}}=0.5~\text{nm}$, $\lambda_{\mathrm{\sigma}}=0.25~\text{nm}$, $n_{\mathrm{Fourier}}=10$. (This roughness-modulated potential is the same as that shown in Fig.~4(b) of the main paper.)   The results are shown in Fig.~S14 and represent the average of 1000 KMC simulations, each of $10^6$ steps.\\

\noindent The global MSD, averaged over all trajectories and starting positions in the simulations, exhibits ``normal'' diffusive scaling,  \mbox{$\langle r^2(t)\rangle \propto t$}, for all three cases. From the perspective of the global MSD, neither the vdW well nor the heterogeneity in the potential are enough to drive anomalous diffusion: the underlying molecular dynamics remain Brownian. In contrast, if we select only those trajectories that start with a molecule entering the detection region and average over \textit{those} events, subdiffusive behaviour is observed (inset to Fig.~S14). This detector-conditioned MSD, however, reflects the persistence and recurrence of locally-confined trajectories rather than indicating a breakdown of normal diffusion.

\subsubsection*{Statistical spread of $\alpha_{\mathrm{ITD}}$ over different roughness realisations} V\'{a}zquez \textit{et al.}\cite{Vazquez2006} examined the spread of values of $\alpha$ around a value of 1 for ITDs measured for three exemplars of bursty human dynamics: visits of a web portal by a single user, emails sent out by a user, and library loans made by a single individual. They found significant spread in the values of $\alpha$ in each case, as evidenced by the widths of the published histograms, but did not provide variance/standard deviation (nor mean) values. We have therefore digitally extracted the distribution data from screenshots of Figs.~2(e),(f), and (g) of their paper\cite{Vazquez2006}, assumed normality, and fitted a Gaussian to each data set. This, admittedly rough and limited, analysis yielded the following estimates: $\alpha_{\mathrm{web}}=1.06 \pm 0.09$, $\alpha_{\mathrm{email}}=1.07 \pm 0.07$, and $\alpha_{\mathrm{library}}=1.04 \pm 0.11$, where the error bar in each case is $\pm 1\sigma$. (Our estimate of $\alpha_{\mathrm{library}}$ is considerably more unreliable than the others due to the scatter in the source distribution. Compare Figs.~2(e),~(f),~and~(g) of Ref.\cite{Vazquez2006}.)\\  

\noindent A central hypothesis in V\'{a}zquez \textit{et al.}\cite{Vazquez2006} is that the spread in the observed values of $\alpha$ arises from the finite number of events recorded for each user in the dataset. In other words, the scatter is attributed to measurement/fitting limitations for a finite data set rather than to a more fundamental origin, i.e. that the browsing/email/library activity of each individual is characterised not by a single ``universal'' value of $\alpha = 1$ but by user-to-user variation around that value. Indeed, Vazquez~\textit{et~al.}\cite{Vazquez2006} explicitly conclude that exponents characterising human behaviour ``take up discrete values''. (See also Barab\'{a}si\cite{Barabasi2005}, and Oliveira and Barab\'{a}si\cite{Oliveira2005}.)\\

\noindent To assess whether a comparably broad scatter of $\alpha$ values can emerge from our simple ``random walk + heterogeneity'' model, we ran a total of 1600 KMC simulations for a single molecule, each comprising $10^7$ kinetic Monte Carlo steps, with all parameters fixed except for the random seed used to generate the roughness landscape. (All other aspects of the roughness field, i.e. $\lambda_{\mathrm{c}}=0.5~\text{nm}$, $\lambda_{\mathrm{\sigma}}=0.25~\text{nm}$, $\epsilon=0.4~k_{\mathrm{B}}T$, $n_{\mathrm{Fourier}}=10$, were held constant throughout.) The resulting distribution of $\alpha_{\mathrm{ITD}}$ values is shown in Fig.~S14, and has $\langle t_{\mathrm{IP}} \rangle=1.08 \pm 0.09$, where the error bar again represents $\pm~1\sigma$. (A Kolmogorov-Smirnov test suggests that the distribution is unlikely to be strictly Gaussian.) Intriguingly, the scatter across different roughness realisations in our KMC simulation is comparable to that observed empirically by V\'{a}zquez \textit{et al.}\cite{Vazquez2006} for the inter-event time distributions of email communication, web browsing, and library visits. 

\newpage


\begin{figure}
\centering
\includegraphics[width=0.5\linewidth]{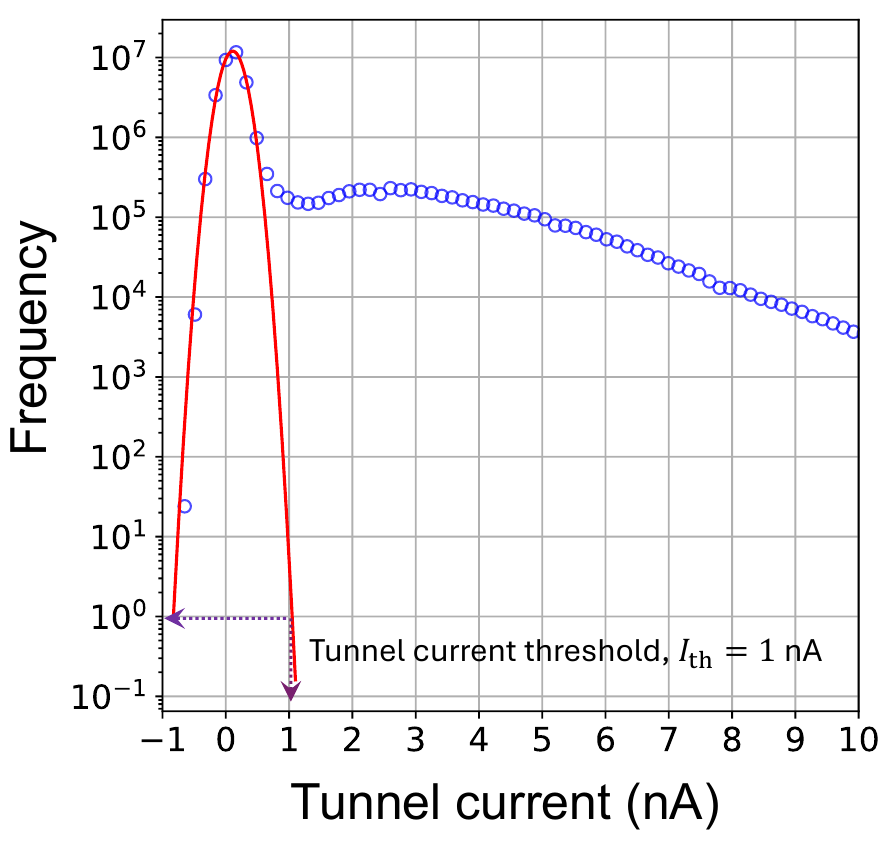}
\caption{\textbf{Selection of the tunnel current threshold for binarisation.} Histogram of tunnel current values acquired for a two-minute time series for PTCDA/Ag(110). (Set-point tunnel current: 200 pA, bias: 500 mV; see Fig. 2(a) (upper plot) and Fig. 3 (upper plot) of the main paper for the corresponding RTD and ITD, respectively.) A Gaussian distribution is fit to the low current values (which arise from instrumental noise). In line with the metric adopted by Hahne \textit{et al.}\cite{Hahne2013}, the threshold current for binarisation of the time series is selected as the point at which the Gaussian fit drops below 1. See also Figs. S2 and S3, and associated discussions in SI Notes \#2 and \#3.}    
\label{SI_Fig1}
\end{figure}

\newpage

\begin{figure}
\centering
\includegraphics[width=0.8\linewidth]{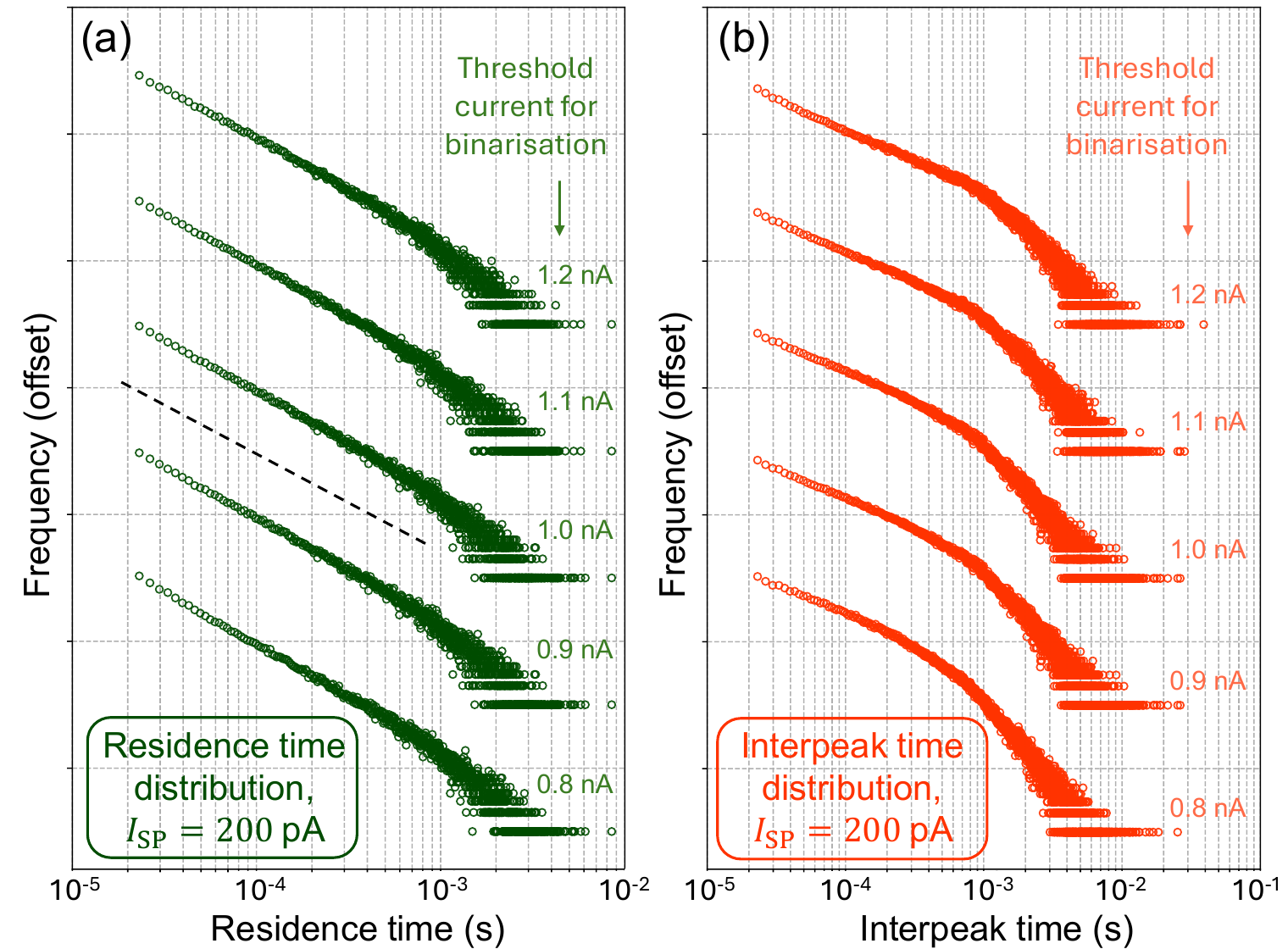}
\caption{\textbf{Effect of varying threshold current for binarisation on RTD/ITD.} \textbf{(a)} Residence time distribution, and \textbf{(b)} interpeak time distribution for a setpoint current of 200 pA where the threshold for binarisation has been varied from 0.8 nA to 1.2 nA, i.e. up to a 20\% change from the value of 1 nA used for the distributions presented in the main paper (see Fig. 2(a), upper curve, and Fig. 3, upper curve.) The dashed line in (a) has a slope of -3/2.}    
\label{SI_Fig2}
\end{figure}

\newpage

\begin{figure}
\centering
\includegraphics[width=0.95\linewidth]{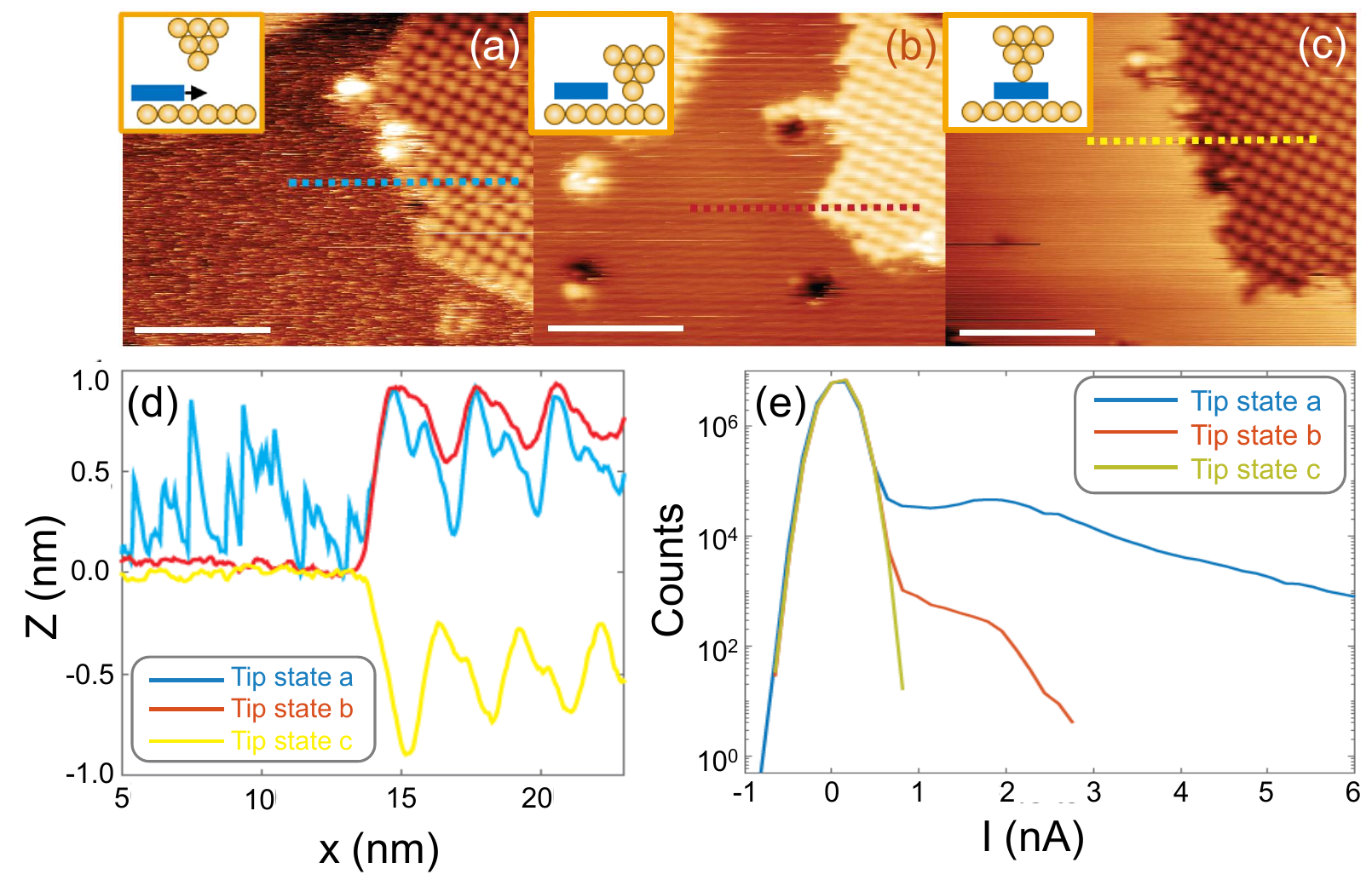}
\caption{\textbf{Influence of tip state.} \textbf{(a), (b), (c)} Constant-current STM images of PTCDA islands (and surrounding 2D molecular gas) on Ag(110) with different tip states, as illustrated schematically by the inset in each case. Image (a) is the same as Fig. 1(a) in the main paper and demonstrates the ideal case for probing diffusion dynamics via tunnel current fluctuations: molecules are free to diffuse under the tip. We interpret the image in (b) (and the corresponding line trace and histogram in (d) and (e), respectively) as arising from a tip that blocks molecular diffusion underneath it but which still produces fluctuations beyond the instrumental noise due to molecular proximity. The contrast inversion in (c) arises from molecular adsorption on the tip\cite{Bohringer1998}; \textbf{(d)} line profiles across the dotted lines shown in (a), (b), and (c); \textbf{(e)} histograms of measured tunnel current values for each of the tip states. Note that tip state c yields only a purely Gaussian distribution arising from instrumental noise as molecular diffusion is entirely blocked; the tip cannot detect diffusing PTCDA molecules.}    
\label{SI_Fig3}
\end{figure}

\newpage

\begin{figure*}[b!]
\centering
\includegraphics[width=0.5\linewidth]{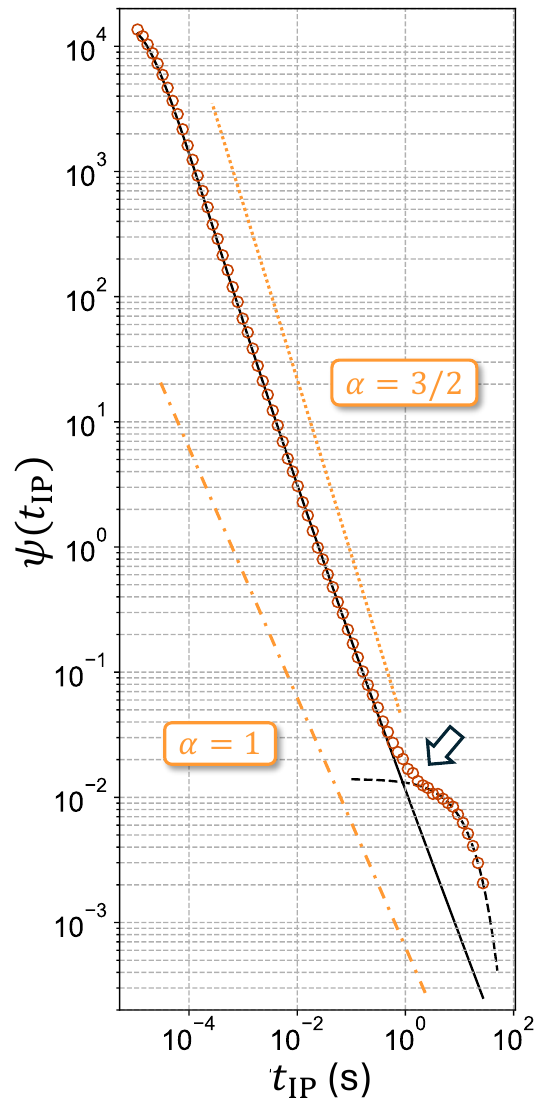}
\caption{\textbf{Comparison of analytical and KMC-generated interpeak-time distributions for unbiased random walk.} \textit{Open circles:} Iinterpeak-time probability density function generated from a kinetic Monte Carlo simulation of a single molecule ($3 \times 5$ lattice sites in area) hopping isotropically ($D_{110}=D_{100}=5000 \text{~nm}^2\text{s}^{-1}$) on a $1000 \times 1000$ lattice in the absence of any tip potential. Total number of Monte Carlo runs: 1000 iterations of $5 \times 10^6$ KMC steps per iteration. Dotted and dash-dotted lines have gradients of $-\frac{3}{2}$ and $-1$, respectively, and are simply guides to the eye. The solid line is a fit to the analytical form of the ITD (Eqn.~S6), assuming an infinite lattice. The dashed line is a fit to the exponentially decaying tail of the distribution arising from periodic boundary conditions (as discussed in the \textit{Marginal recurrence and finite size effects} section of the Supplementary Text.) Its onset is highlighted by the arrow. Note that we have used logarithmic binning for the distribution in order to accentuate the exponential tail.} 
\label{SI_Fig_4}
\end{figure*}

\newpage

\begin{figure}
\centering
\includegraphics[width=0.99\linewidth]{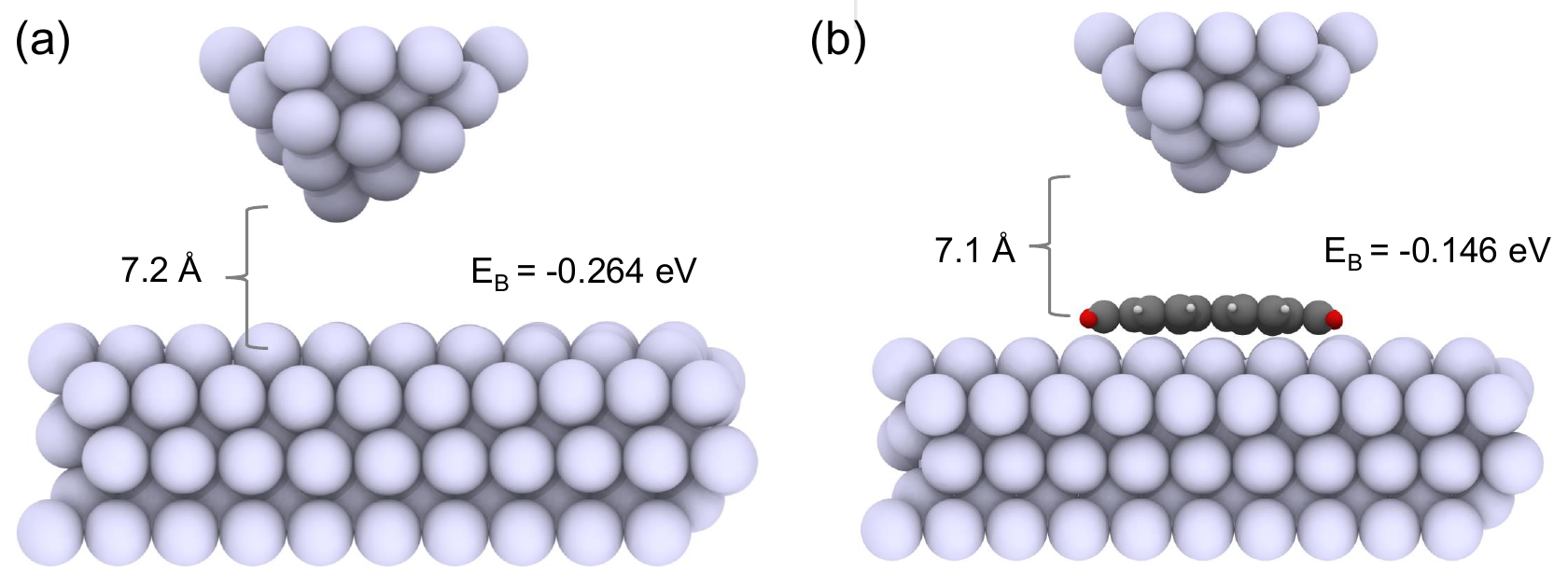}
\caption{\textbf{Converged VASP-PAW-D2 geometries} and binding energies for \textbf{(a)} a silver tip above a Ag(110) surface, and \textbf{(b)} a silver tip above a PTCDA molecule adsorbed on Ag(110), both at a separation of $\sim$ 7\AA. The uppermost two layers of the tip are pinned during the relaxation process. Note that 0.146 eV $\sim~6~k_{\mathrm{B}}T$ at room temperature.}   
\label{SI_Fig5}
\end{figure}

\begin{figure}
\centering
\includegraphics[width=0.99\linewidth]{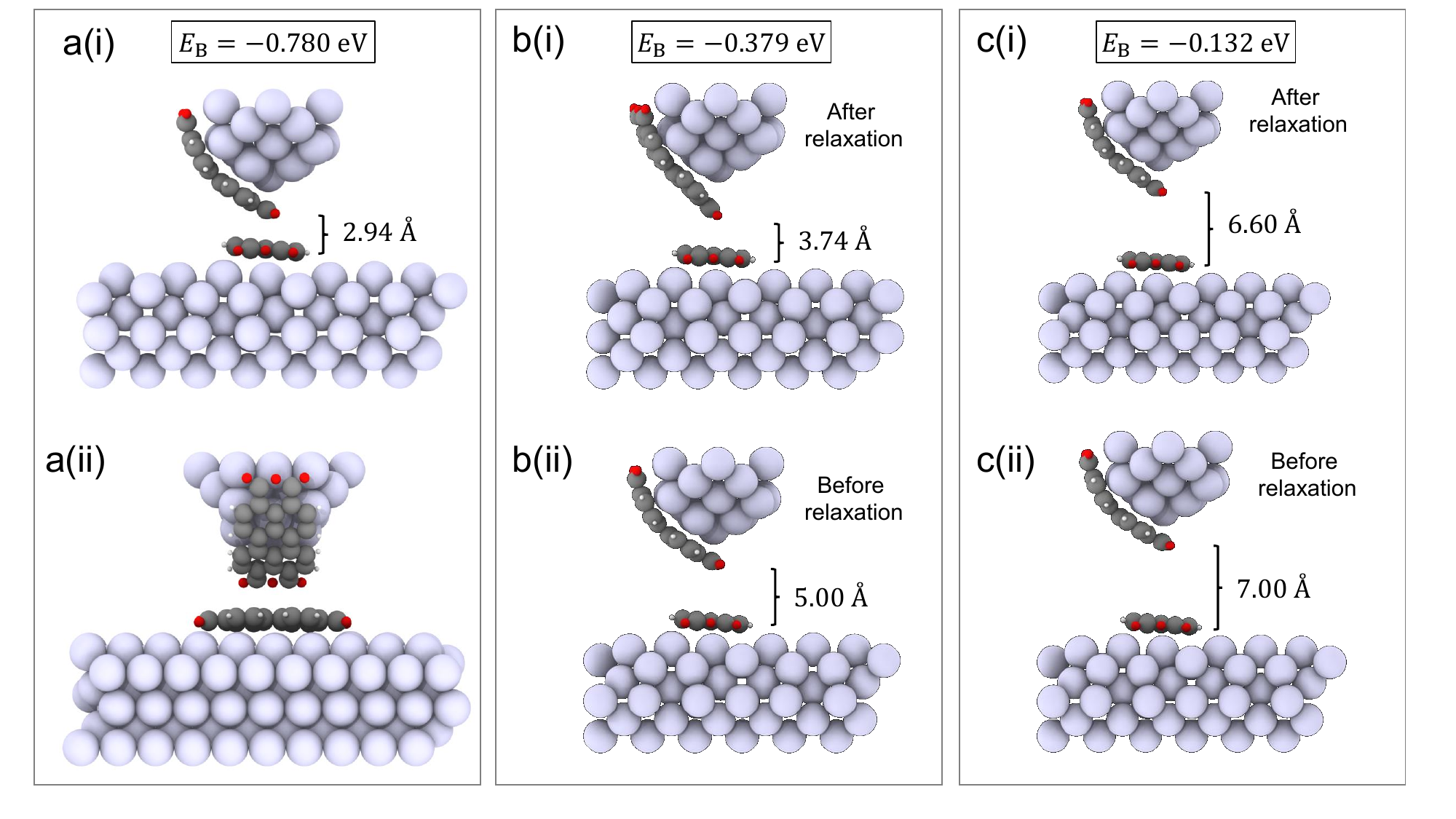}
\caption{\textbf{``Pointed'' PTCDA tip termination.} Converged geometries resulting from VASP-PAW-D2 density functional theory (DFT) calculations. As for Fig. S5, the uppermost two layers of the tip structure are pinned in place (i.e. not free to relax). \textbf{(a)(i)} ``Side-on'' and \textbf{(a)(ii)} ``face-on'' views of a PTCDA-terminated Ag tip at a tip-sample separation of~$\sim 3$~\AA. The binding energy, $E_{\mathrm{B}}$ in this case is -0.78 eV. However, a $\sim 3$~\AA~separation is much smaller than expected for the STM operating conditions in our experiment. We therefore also calculated $E_{\mathrm{B}}$ for \textbf{(b)(ii)}, a pre-relaxation tip-sample separation of $5$~\AA , and \textbf{(c)(ii)} a pre-relaxation separation of $\sim 7$~\AA, which is more in line with the tip-sample gap expected under the relatively low tunnel current setpoint conditions used in our experiments. \textbf{(b)(i)} and \textbf{(c)(i)} show the relaxed geometries. We adopted the value of $E_{\mathrm{B}}$ for the latter as an initial baseline estimate of the depth of the tip-sample interaction potential in the KMC simulations. We again stress, however, that the precise form of the tip–sample interaction potential is not of primary importance. Instead, the key requirement is that the energy landscape exhibits sufficient spatial heterogeneity to generate a broad distribution of return times and thus capture the $\alpha_{\mathrm{ITD}}$ scaling (while retaining the $\sim 3/2$ short-time scaling of the residence time distribution.). The tip potential used in the KMC simulations, and informed by the DFT calculations presented here, serves as a physically motivated model that captures the relevant interaction strength and reproduces the characteristic timescales observed experimentally.}    
\label{SI_Fig6}
\end{figure}

\begin{figure}
\centering
\includegraphics[width=0.99\linewidth]{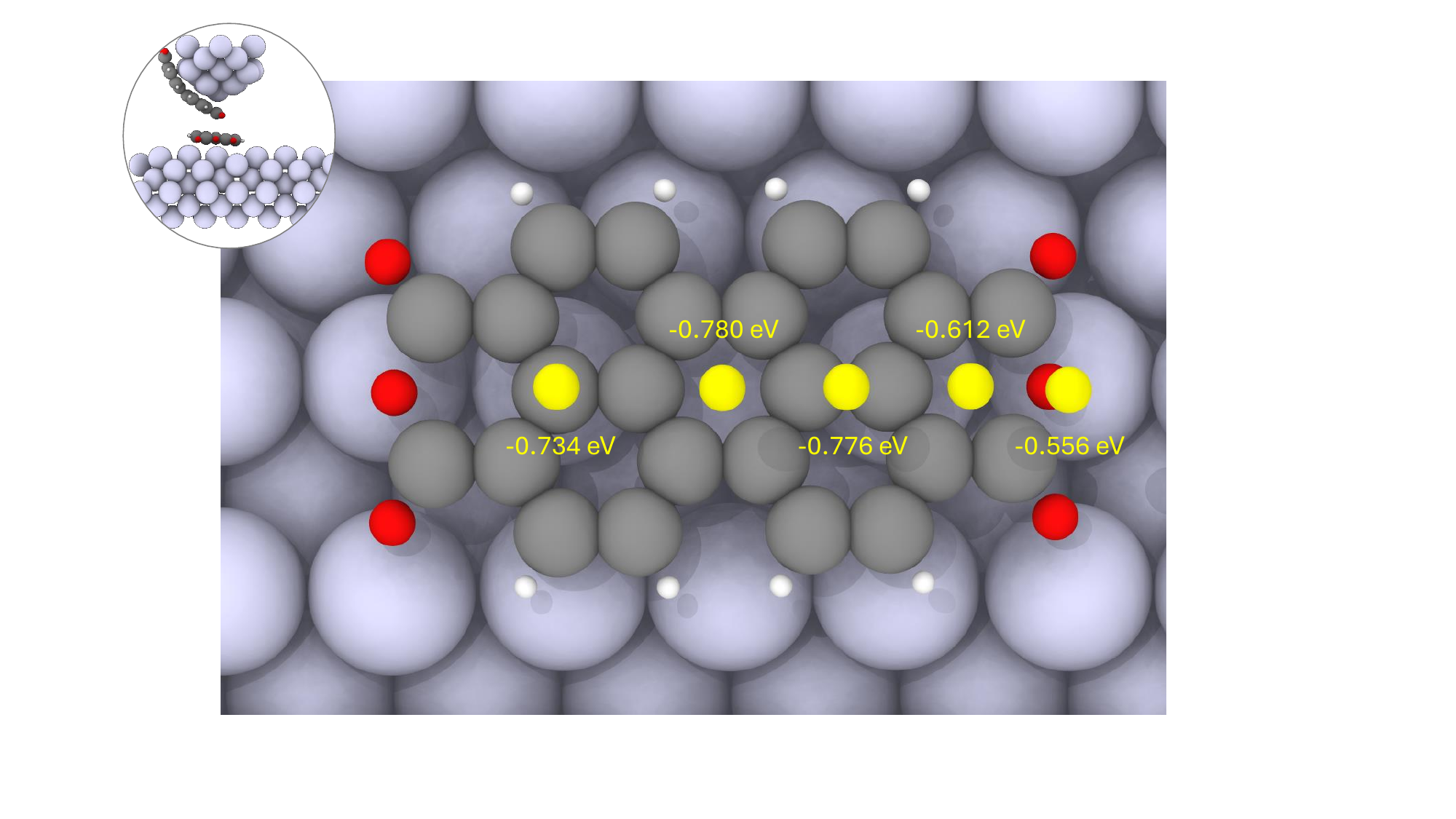}
\caption{\textbf{Intramolecular variation of binding energy} The ``pointed'' PTCDA tip of Fig. S6 (also shown in the inset, upper left), at a tip-sample separation of $\sim$~3~\AA~ was placed at a variety of intramolecular sites for the surface-adsorbed Ag(110). There is an appreciable variation of $E_{\mathrm{B}}$ across the molecule, providing another mechanism for heterogeneity in the energy landscape.}    
\label{SI_Fig7}
\end{figure}

\begin{figure}
\centering
\includegraphics[width=0.99\linewidth]{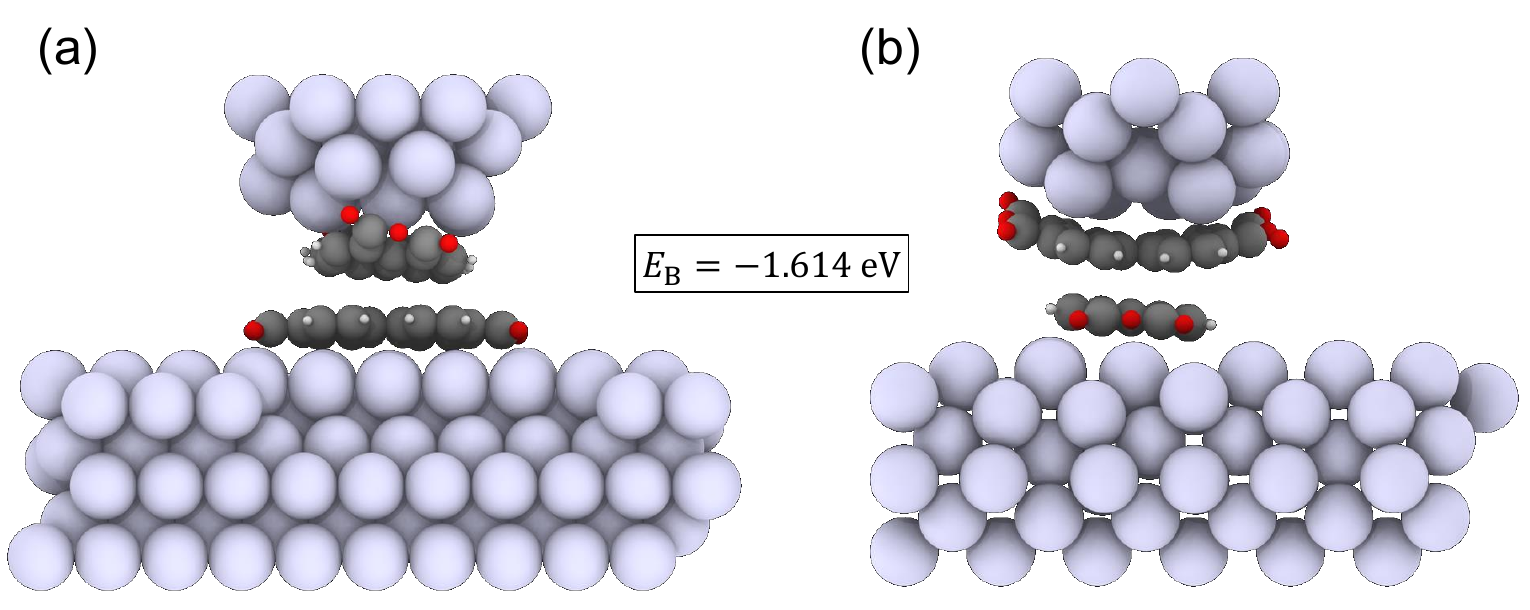}
\caption{\textbf{``quasi-planar'' PTCDA tip} \textbf{(a)}, \textbf{(b)} ``face-on'' and ``side-on'' views of a PTCDA-functionalised tip (at a tip-sample separation of $\sim 3$~\AA),~ where the molecule adopts a relatively planar (as compared to Figs. S5 and S6), albeit bent, adsorption geometry. We speculate that a termination of this type may underpin the absence of diffusive signal seen for tip state C in Fig. S3. See also Movie S1.}    
\label{SI_Fig8}
\end{figure}

\newpage

\begin{figure}
\centering
\includegraphics[width=0.7\linewidth]{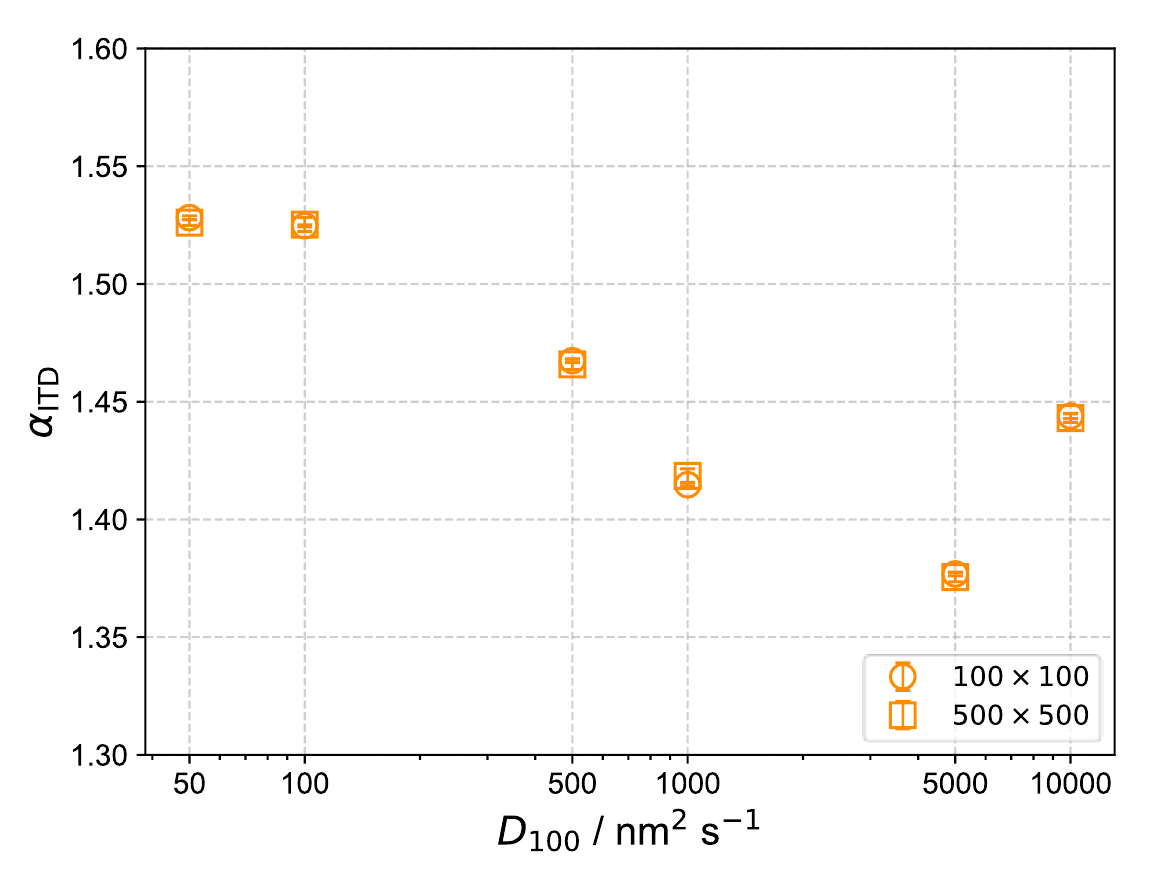}
\caption{\textbf{Anisotropy is not sufficient to yield $\alpha_{\mathrm{ITD}}\sim 1$.} Scaling exponent, $\alpha_{\mathrm{ITD}}$ determined via maximum likelihood estimation (in the $20~\mu\text{s} \leq t_{\mathrm{IP}} \leq 1~\text{ms}$ range) from a series of KMC simulations with a flat-bottomed van der Waals well of depth $6~k_{\mathrm{B}}T$, no roughness field, a single molecule, a fixed value of $D_{110}=5000~\text{nm}^2\text{s}^{-1}$, and with the diffusion coefficient in the orthogonal direction, $D_{100}=50, 100, 500, 1000, 5000, 10000~\text{nm}^2\text{s}^{-1}.$ Ten simulations were run for each value of $D_{100}$ and the standard error in the mean taken as the uncertainty in the value of $\alpha_{\mathrm{ITD}}$. The error bars are, however, smaller than the size of the data points. \textit{Open circles:} $100 \times 100$ lattice. \textit{Open squares:}$500 \times 500$ lattice.}     
\label{SI_Fig9}
\end{figure}
\newpage

\begin{figure}
\centering
\includegraphics[width=0.95\linewidth]{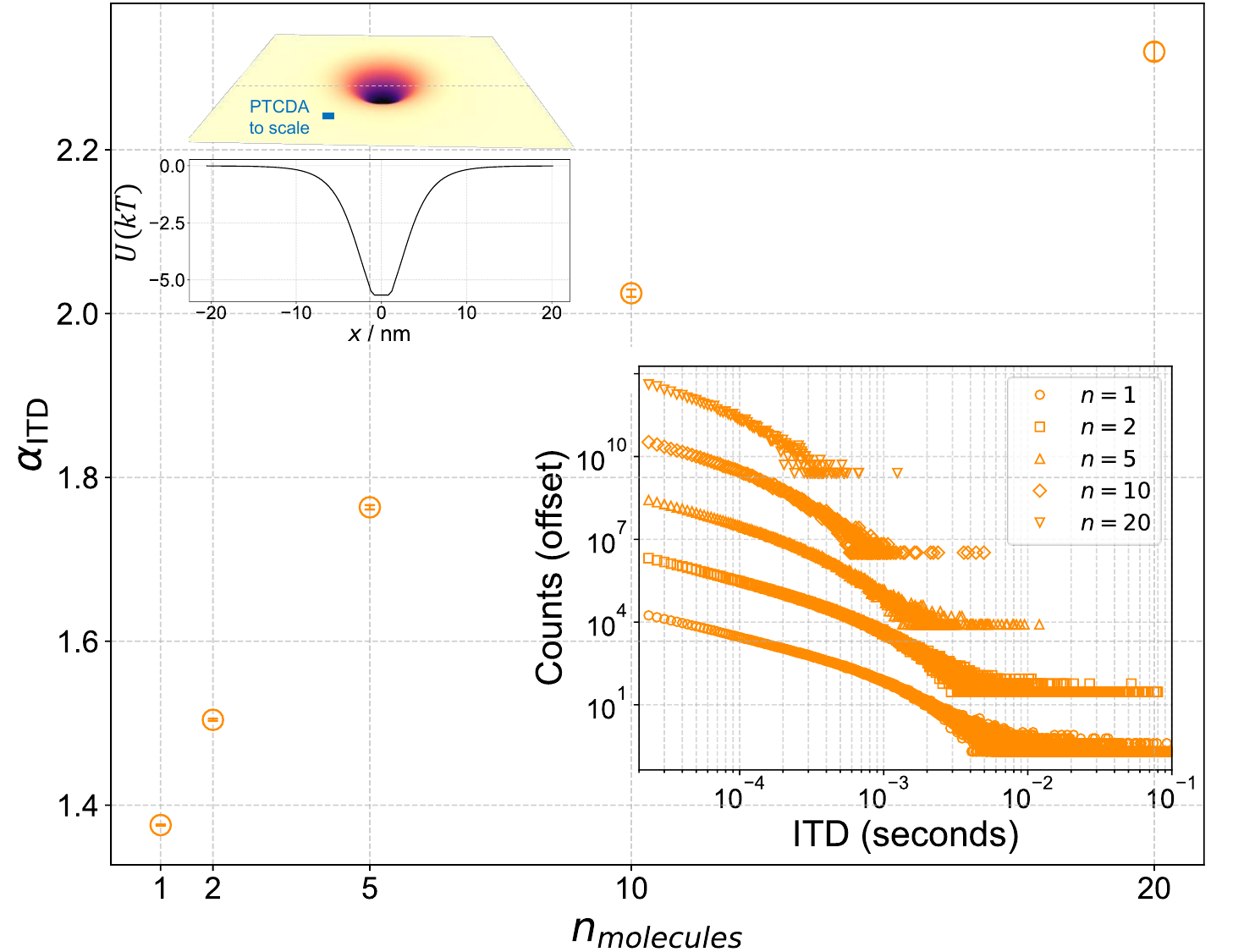}
\caption{\textbf{Increasing molecular number density does not yield $\alpha_{\mathrm{ITD}} \sim 1$.} Scaling exponent, $\alpha_{\mathrm{ITD}}$, determined via maximum likelihood estimation (in the \mbox{$20~\mu\text{s} \leq t_{\mathrm{IP}} \leq 1~\text{ms}$} range) from a series of KMC simulations with a flat-bottomed van der Waals well of depth $6~k_{\mathrm{B}}T$ and no roughness field (see inset to left), $D_{110}=D_{100}=5000~\text{nm}^2\text{s}^{-1}$, and an increasing number of simulated molecules, $n_{\mathrm{molecules}}=1,2,5,10,20$.  Ten simulations were run for each value of $n_{\mathrm{molecules}}$ and the standard error in the mean taken as the uncertainty in the value of $\alpha_{\mathrm{ITD}}$. The error bars are, however, smaller than the size of the data points. Lattice size=$100 \times 100$. \textit{Inset,~lower~right:} Aggregated ITDs (over ten KMC runs, each of $10^7$ steps) as a function of $n$ molecules. The apparent increase in $\alpha_{\mathrm{ITD}}$ as a function of $n$ arises simply because the exponential tail of the distribution shifts to shorter times as the molecular number density increases.}     
\label{SI_Fig10}
\end{figure}

\begin{figure}
\centering
\includegraphics[width=0.95\linewidth]{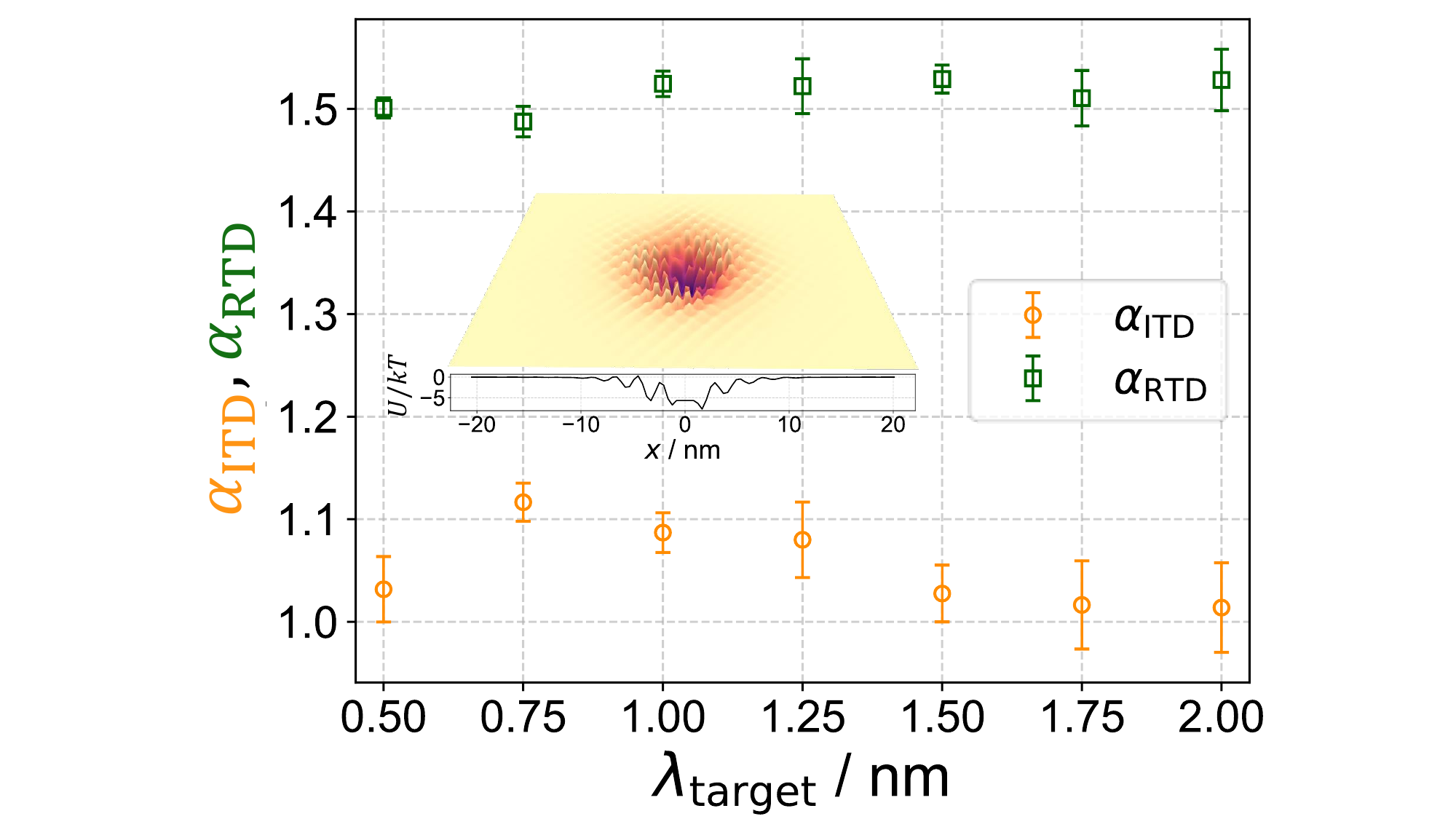}
\caption{\textbf{Variation of $\alpha_{\mathrm{RTD}}$ and $\alpha_{\mathrm{ITD}}$ with $0.5~\text{nm}\leq \lambda_{\mathrm{c}}\leq2.0~\text{nm}$.} Each $\lambda_{\mathrm{c}}$ data-point is the mean value determined from ten KMC simulations, each of $10^7$~steps, and each with a different random seed for the roughness field generation. KMC parameters: $n_{\mathrm{molecules}}=1$, $D_{110}=D_{100}=5000~\text{nm}^2\text{s}^{-1}$, effective detection radius: 0.91~$\text{nm}$, radius of curvature of simulated tip: 30 nm, depth of vdW well: $6~k_{\mathrm{B}}T$, number of Fourier modes for roughness field:1, $\lambda_{\mathrm{\sigma}}=0.25~\text{nm}$ (i.e. the standard deviation for the random choice of wavelength centred on $\lambda_{\mathrm{c}}$), $\epsilon=0.4~k_{\mathrm{B}}T$. The error bars are the standard error in the mean across the ten KMC simulations. \textbf{Inset:}~pseudo-3D plot and 1D profile (through the centre of the well) of one realisation of the potential.}
\label{SI_Fig11}
\end{figure}

\begin{figure}
\centering
\includegraphics[width=0.95\linewidth]{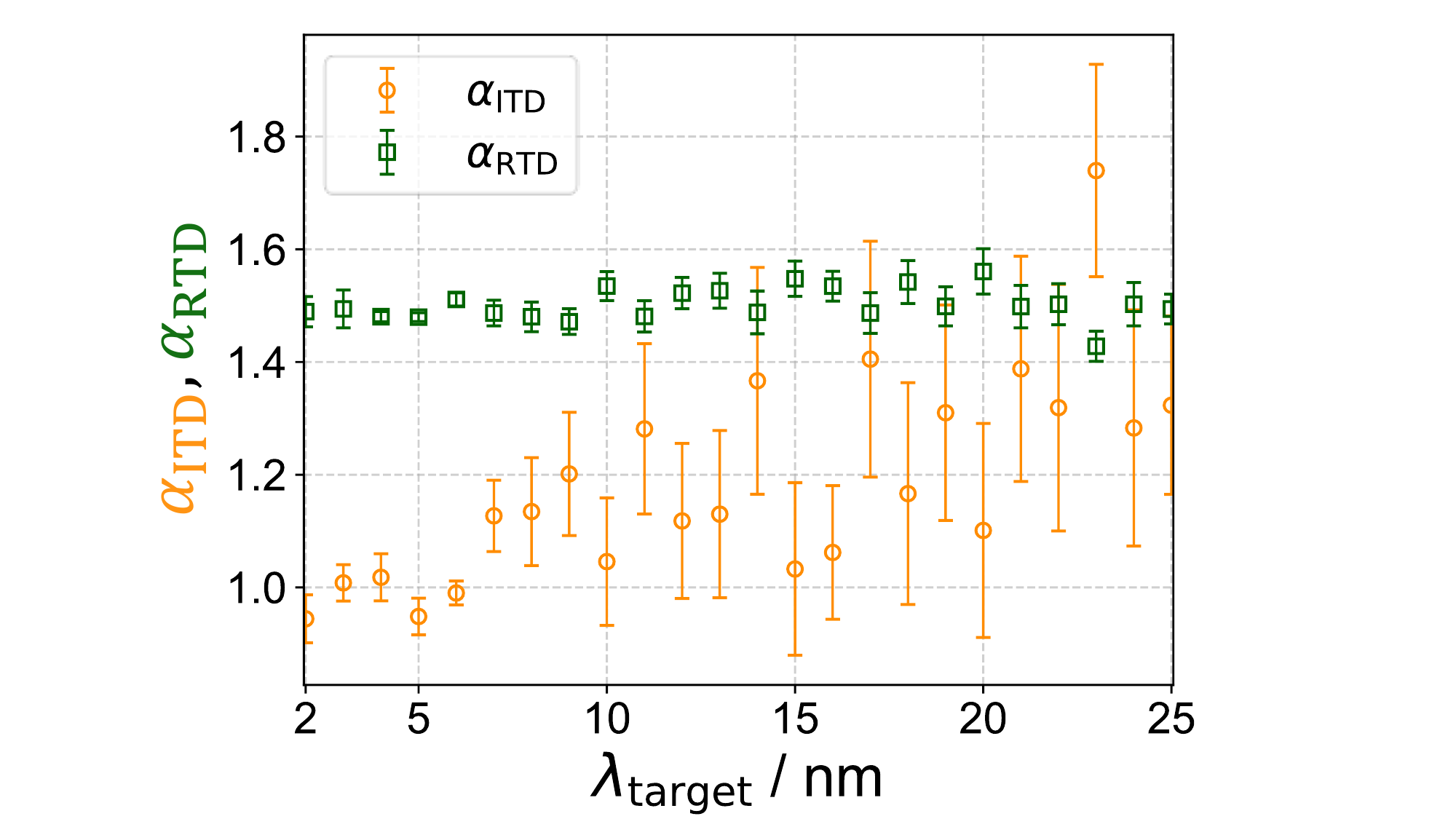}
\caption{\textbf{Variation of $\alpha_{\mathrm{RTD}}$ and $\alpha_{\mathrm{ITD}}$ with $2~\text{nm}\leq \lambda_{\mathrm{c}}\leq25~\text{nm}$.} As for Fig. S11, each $\lambda_{\mathrm{c}}$ data-point is the mean value determined from ten KMC simulations, each of $10^7$~steps, and each with a different random seed for the roughness field generation. KMC parameters: $n_{\mathrm{molecules}}=1$, $D_{110}=D_{100}=5000~\text{nm}^2\text{s}^{-1}$, effective detection radius: 0.91~$\text{nm}$, radius of curvature of simulated tip: 30 nm, depth of vdW well: $6~k_{\mathrm{B}}T$, number of Fourier modes for roughness field:1, $\lambda_{\mathrm{\sigma}}=0.25~\text{nm}$ (i.e. the standard deviation for the random choice of wavelength centred on $\lambda_{\mathrm{c}}$), $\epsilon=0.4~k_{\mathrm{B}}T$. The error bars are the standard error in the mean across ten KMC simulations.}
\label{SI_Fig12}
\end{figure}

\begin{figure}
\centering
\includegraphics[width=0.95\linewidth]{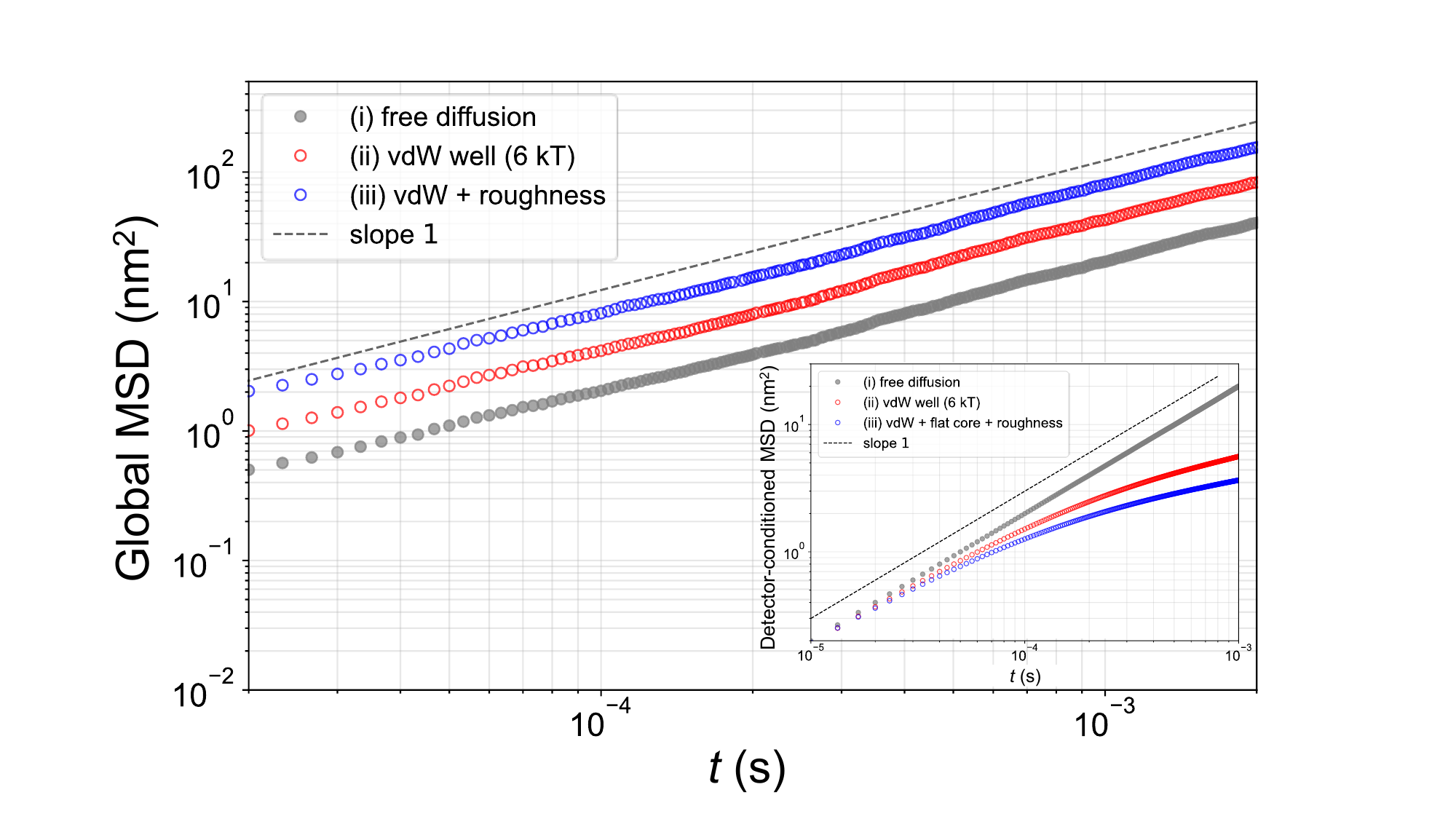}
\caption{\textbf{Mean square displacement (MSD) in range $20~\mu\text{s}\leq t \leq 2~\text{ms}$ for (i)} free diffusion -- no tip potential; \textbf{(ii)} van der Waals (vdW) well of depth 6$kT$ but with no roughness field; \textbf{(iii)} vdW well of depth 6$kT$ with roughness field ($\epsilon=0.4~kT$, $n_{\mathrm{Fourier}}=10$, $\lambda_{\mathrm{c}}=0.5~\text{nm}$, $\lambda_{\mathrm{\sigma}}=0.25~\text{nm}$). \textbf{Inset:} ``detector-conditioned'' MSD under identical conditions and over the same temporal range. For the ``detector-conditioned'' MSD we select for trajectories that begin at times when the walker enters the detection region and then average $\langle|\mathbf{r}(t+\tau)-\mathbf{r}(t)|^2\rangle$ over that subset of events. The global MSD exhibits uniform diffusive behaviour for all three conditions, i.e. $\langle r^2(t)\rangle \propto t$. Only the detector-conditioned (and therefore statistically biased) MSD shows ``subdiffusive'' behaviour. The MSD plots are the result of averaging over 1000~KMC~simulations, each of $10^6$ steps. The dashed line is simply a guide to the eye to highlight the gradient of 1 expected for normal diffusion.}  
\label{SI_Fig13}
\end{figure}

\begin{figure}
\centering
\includegraphics[width=0.95\linewidth]{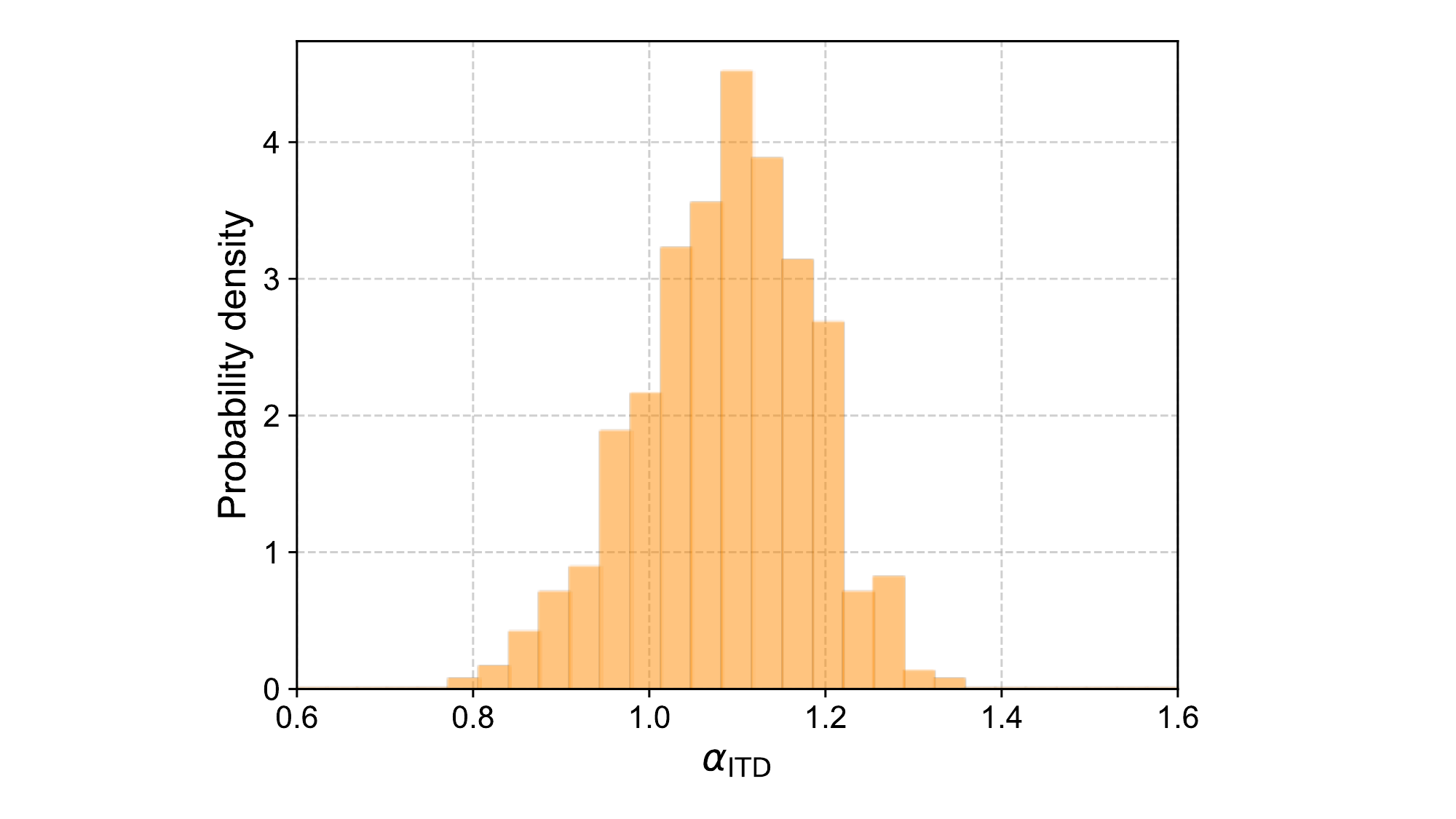}
\caption{\textbf{Statistical spread of $\alpha_{\mathrm{ITD}}$ over different roughness realisations.} The histogram shows the scatter of $\alpha_{\mathrm{ITD}}$ values across 1600 KMC runs \mbox{($\langle \alpha_{\mathrm{ITD}} \rangle=1.08,~\sigma=0.09$)}, each of $10^7$ steps, with the following parameters: one molecule; $100 \times 100$ lattice;  $D_{110}=D_{100}=5000~\text{nm}^2\text{s}^{-1}$; flat-bottomed vdW well of depth $6~k_{\mathrm{B}}T$ and a roughness field with $\epsilon=0.4~k_{\mathrm{B}}T$, $\lambda_{\mathrm{c}}=0.5~\text{nm}$, $\lambda_{\mathrm{\sigma}}=0.25~\text{nm}$, $n_{\mathrm{Fourier}}=10$. }
\label{SI_Fig14}
\end{figure}

\clearpage


\begin{table*}[h!]
\centering
\renewcommand{\arraystretch}{0.6}
\begin{tabular}{clccc}
\hline
$I_{\mathrm{SP}}$ (pA) & Model & $n_{\mathrm{params}}$ & $\Delta$AIC & $\Delta$BIC \\
\hline
\multirow{6}{*}{200}
& Exponentially truncated power law ($\alpha, t_c$, $\beta=1$) & 2 & 0 & 0 \\
& KWW-tempered power law ($\alpha, t_c, \beta$) & 3 & 0 & 10 \\
& Weibull (KWW) & 2 & 2{,}157 & 2{,}157 \\
& Mixture of two exponentials & 3 & 7{,}098 & 7{,}108 \\
& Lognormal ($\mu, \sigma$) & 2 & 9{,}201 & 9{,}201 \\
& Power law ($\alpha$) & 1 & 49{,}554 & 49{,}544 \\
\hline
\multirow{6}{*}{150}
& KWW-tempered power law ($\alpha, t_c, \beta$) & 3 & 0 & 0 \\
& Exponentially truncated power law ($\alpha, t_c$, $\beta=1$) & 2 & 78 & 68 \\
& Weibull & 2 & 2{,}952 & 2{,}942 \\
& Mixture of two exponentials & 3 & 6{,}431 & 6{,}431 \\
& Lognormal ($\mu, \sigma$) & 2 & 9{,}684 & 9{,}674 \\
& Power law ($\alpha$) & 1 & 45{,}826 & 45{,}806 \\
\hline
\multirow{6}{*}{125}
& KWW-tempered power law ($\alpha, t_c, \beta$) & 3 & 0 & 0 \\
& Exponentially truncated power law ($\alpha, t_c$, $\beta=1$) & 2 & 184 & 174 \\
& Weibull & 2 & 3{,}203 & 3{,}193 \\
& Mixture of two exponentials & 3 & 5{,}642 & 5{,}642 \\
& Lognormal ($\mu, \sigma$) & 2 & 9{,}869 & 9{,}859 \\
& Power law ($\alpha$) & 1 & 38{,}322 & 38{,}302 \\
\hline
\multirow{6}{*}{100}
& KWW-tempered power law ($\alpha, t_c, \beta$) & 3 & 0 & 0 \\
& Exponentially truncated power law ($\alpha, t_c$, $\beta=1$) & 2 & 558 & 548 \\
& Weibull & 2 & 4{,}211 & 4{,}201 \\
& Mixture of two exponentials & 3 & 4{,}786 & 4{,}786 \\
& Lognormal ($\mu, \sigma$) & 2 & 9{,}277 & 9{,}267 \\
& Power law ($\alpha$) & 1 & 25{,}413 & 25{,}393 \\
\hline
\multirow{6}{*}{75}
& KWW-tempered power law ($\alpha, t_c, \beta$) & 3 & 0 & 0 \\
& Exponentially truncated power law ($\alpha, t_c$, $\beta=1$) & 2 & 809 & 799 \\
& Mixture of two exponentials & 3 & 4{,}093 & 4{,}093 \\
& Weibull & 2 & 4{,}464 & 4{,}455 \\
& Lognormal ($\mu, \sigma$) & 2 & 8{,}865 & 8{,}856 \\
& Power law ($\alpha$) & 1 & 19{,}853 & 19{,}835 \\
\hline
\multirow{6}{*}{50}
& KWW-tempered power law ($\alpha, t_c, \beta$) & 3 & 0 & 0 \\
& Exponentially truncated power law ($\alpha, t_c$, $\beta=1$) & 2 & 940 & 931 \\
& Mixture of two exponentials & 3 & 2{,}590 & 2{,}590 \\
& Weibull & 2 & 4{,}069 & 4{,}060 \\
& Lognormal ($\mu, \sigma$) & 2 & 6{,}395 & 6{,}386 \\
& Power law ($\alpha$) & 1 & 10{,}082 & 10{,}065 \\
\hline
\multirow{6}{*}{KMC (Fig. 4(a))}
& KWW-tempered power law ($\alpha, t_c, \beta$) & 3 & 0 & 0 \\
& Exponentially truncated power law ($\alpha, t_c$, $\beta=1$) & 2 & 9{,}736 & 9{,}724 \\
& Weibull & 2 & 11{,}231 & 11{,}219 \\
& Lognormal ($\mu, \sigma$) & 2 & 58{,}172 & 58{,}160 \\
& Mixture of two exponentials & 3 & 151{,}207 & 151{,}207 \\
& Power law ($\alpha$) & 1 & 387{,}723 & 387{,}698 \\
\hline

\end{tabular}
\renewcommand{\arraystretch}{1.0}

\caption{Information-criterion comparison of candidate models for the measured interpeak-time distributions at all setpoints and for a representative KMC-generated ITD (see Fig. 4(a) of main paper). Lower $\Delta$AIC and $\Delta$BIC indicate stronger support relative to the best model for each dataset. Each value has been quoted to the nearest integer.}
\label{tab:ic_all}
\end{table*}


\clearpage 

\paragraph{Caption for Movie S1.}
{DFT simulation of ``quasi-flat'' PTCDA termination driving motion of surface-adsorbed PTCDA towards the tip.}



\end{document}